\documentclass[twocolumn]{aastex631}
\usepackage{amsmath}
\usepackage{mathtools}
\usepackage{amssymb}

\shorttitle{Precessing magnetars in SLSNe}
\shortauthors{Zhang et al}

\begin{document}
\title{Hydrogen-poor Superluminous Supernovae with Bumpy Light Curves Powered by Precessing Magnetars}

\author[0009-0002-0930-1830]{Biao Zhang}
\affiliation{Department of Astronomy, University of Science and Technology of China, Hefei 230026, People’s Republic of China; daizg@ustc.edu.cn}
\affiliation{School of Astronomy and Space Science, University of Science and Technology of China, Hefei 230026, People’s Republic of China}

\author[0000-0002-8391-5980]{Long Li}
\affiliation{Department of Physics, School of Physics and Materials Science, Nanchang University, Nanchang 330031, People’s Republic of China}

\author[0000-0002-7835-8585]{Zi-Gao Dai}
\affiliation{Department of Astronomy, University of Science and Technology of China, Hefei 230026, People’s Republic of China; daizg@ustc.edu.cn}
\affiliation{School of Astronomy and Space Science, University of Science and Technology of China, Hefei 230026, People’s Republic of China}

\author[0000-0002-1766-6947]{Shu-Qing Zhong}
\affiliation{School of Science, Guangxi University of Science and Technology, Liuzhou 545006, People’s Republic of China}
\affiliation{Department of Astronomy, University of Science and Technology of China, Hefei 230026, People’s Republic of China; daizg@ustc.edu.cn}
\affiliation{School of Astronomy and Space Science, University of Science and Technology of China, Hefei 230026, People’s Republic of China}

\begin{abstract}
Recent observations and statistical studies have revealed that a significant fraction of hydrogen-poor superluminous supernovae (SLSNe-I) exhibit light curves that deviate from the smooth evolution predicted by the magnetar-powered model, instead showing one or more bumps after the primary peak. However, the formation mechanisms of these post-peak bumps remain a matter of debate. Furthermore, previous studies employing the magnetar-powered model have typically assumed a fixed magnetic inclination angle and neglected the effects of magnetar precession. However, recent research has shown that the precession of newborn magnetars forming during the collapse of massive stars causes the magnetic inclination angle to evolve over time, thereby influencing magnetic dipole radiation. In this paper, therefore, we incorporate the effects of magnetar precession into the magnetar-powered model to develop the precessing magnetar-powered model. Using this model, we successfully reproduce the multi-band light curves of 6 selected representative SLSNe-I with post-peak bumps. Moreover, the derived model parameters fall within the typical parameter range for SLSNe-I. By combining the precessing magnetars in SLSNe-I and long GRBs, we find that the ellipticity of magnetars is related to the dipole magnetic field strength, which may suggest a common origin for the two phenomena. Our work provides a potential explanation for the origin of post-peak bumps in SLSNe-I and offers evidence for the early precession of newborn magnetars formed in supernova explosions.
\end{abstract}

\section{Introduction}
\label{sec:intro}
Superluminous supernovae (SLSNe) are a special class of supernovae observed within the past twenty years, characterized by extremely high luminosities, reaching tens to hundreds of times that of normal supernovae \citep{Gal-Yam2012,Gal-Yam2019}. From the perspective of spectral features, SLSNe include hydrogen-poor SLSNe-I and hydrogen-rich SLSNe-II types. The radioactive decay of $^{56}\mathrm{Ni}$ is one of the important energy sources for normal supernovae, but this mechanism is not applicable to most SLSNe-I and SLSNe-II. This is because the high luminosity of SLSNe-I and SLSNe-II requires a much larger amount of $^{56}\mathrm{Ni}$ than can be synthesized in a regular supernova explosion \citep{Umeda&Nomoto2008,WangShan-Qin2019}.

The energy source for SLSNe-II is typically attributed to the interaction between the ejecta produced during the supernova explosion and the circumstellar medium (CSM) formed by the progenitor prior to the explosion \citep{Smith&McCray2007,Chevalier&Irwin2011,Chatzopoulos2012}. The interaction between the ejecta and the CSM is also one of the primary energy sources for SLSNe-I \citep{Chatzopoulos2012,Ginzburg&Balberg2012,Inserra2017,LiLong2020}. However, the difference is that for SLSNe-I, the CSM is hydrogen-poor, whereas in SLSNe-II, the CSM is hydrogen-rich. Another primary energy source model for SLSNe-I is the magnetar central engine model \citep{Kasen&Bildsten2010,Woosley2010,Dessart2012,Metzger2015,WangShan-Qin2015}. In this scenario, the magnetar formed after the supernova explosion injects its energy into the supernova ejecta via magnetic dipole radiation, thereby producing the light curve of the SLSNe. Both the CSM interaction model and the magnetar central engine model can reproduce the observed light curves of some SLSNe-I \citep{Chatzopoulos2013,LiuLiang-Duan2017,Nicholl2017,YuYun-Wei2017}. Further investigation is needed to distinguish between the two models, and studying the bumps in the light curves of SLSNe-I may provide some clues.

\cite{Hosseinzadeh2022} and \cite{Chen2023} conducted systematic studies on the properties of light curves of two different hydrogen-poor SLSNe-I samples, finding that the light curves of many hydrogen-poor SLSNe-I are not smooth at all epochs and often display one or more post-peak bumps, observed in approximately 30\% to 80\% of the total sample. The results of the studies by \cite{Hosseinzadeh2022} and \cite{Chen2023} suggest that post-peak bumps are relatively common in the light curves of hydrogen-poor SLSNe-I. The origin of these post-peak bumps may be linked to the energy sources of SLSNe-I. The single CSM interaction and the original magnetar central engine model can only account for the observations of SLSNe-I with smooth light curves and have difficulty explaining the observations of SLSNe-I with post-peak bumps. \cite{Hosseinzadeh2022} and \cite{Chen2023} suggest that multiple CSM interactions or variations in the energy injection from the central engine may provide an explanation for the origin of post-peak bumps in SLSNe-I. In addition, there exists another subclass of SLSNe-I whose light curves exhibit a double-peaked structure, with a dimmer peak appearing prior to the primary peak. \cite{Kasen2016} proposed a magnetar-driven shock breakout model to explain such SLSNe-I.

Some studies have proposed different models in an attempt to explain the formation of post-peak bumps in SLSNe-I. \cite{LiLong2020} used double CSM interactions to explain post-peak bumps of PS1-12cil, while \cite{LiuLiang-Duan2018} employed triple CSM interactions to account for post-peak bumps of iPTF15esb. \cite{YuYun-Wei&Li2017} and \cite{DongXiao-Fei2023} suggest that post-peak bumps in SLSNe, such as SN 2015bn, SN 2018kyt, and SN 2019stc, may be produced by flare activity of a magnetar. \cite{Moriya2022} attempt to reproduce the post-peak bumps of SN 2015bn and SN 2019stc by considering the time-dependent thermalization efficiency of the energy injected by the magnetar. \cite{ZhuJin-Ping2024} propose a magnetar-star binary engine to interpret the observations of several SLSNe-I with a single post-peak bump. Although the various models mentioned above attempt to explain the origin of post-peak bumps in SLSNe-I, the mechanisms behind their formation remain debated. Several points regarding these models are worth noting: the lack of narrow emission lines in the spectra of most SLSNe-I poses a challenge to the CSM interaction model; the formation mechanism of magnetar flares and the physical processes driving variations in the thermalization efficiency of injected energy in SLSNe are not yet fully understood; and the magnetar-star binary engine model appears more suited to explaining a single post-peak bump. Therefore, further research is required to uncover the physical origin of post-peak bumps in SLSNe-I.

The energy injected by a magnetar into the supernova ejecta is related to the magnetic inclination angle, defined as the angle between the rotation axis and the magnetic axis. For a deformed magnetar deviating from a spherical shape, the magnetic inclination angle evolves over time due to the precession of the magnetar \citep{Zanazzi&Lai2015,ZhangBiao2024}. Consequently, the energy injected into the supernova ejecta by a precessing magnetar varies over time, which could potentially lead to the appearance of bumps in the light curves of SLSNe. Currently, several studies have provided evidence for magnetar precession. For instance, some works propose precessing magnetars as the central engine of gamma-ray bursts (GRBs) to explain the quasi-periodic oscillation (QPO) signals observed in the X-ray afterglow plateaus \citep{Suvorov&Kokkotas2020,Zou&Liang2022,ZhangBiao2024}. Additionally, the periodic phase modulation in the hard X-ray pulses of 4U 0142+61 has been considered evidence for magnetar precession \citep{Makishima2014}. Furthermore, the periodicity of FRB 180916 has been attributed to magnetar precession in some studies \citep{Levin2020,Zanazzi&Lai2020,WeiYu-Jia2022}.

The purpose of this work is to propose a precessing deformed magnetar as the central engine of SLSNe and to explain the formation of post-peak bumps in some well-observed SLSNe-I by considering the energy variation injected into the supernova ejecta caused by the evolution of the magnetic inclination angle during the precession of the deformed magnetar. The contents of this paper are organized as follows. The precessing magnetar central engine model is introduced in Section \ref{sec:precessing magnetar}. In Section \ref{sec:Lightcurv}, we present the observational properties and light curve fitting results for six SLSNe-I with post-peak bumps. The analysis of the results is presented in Section \ref{sec:resulst}. Finally, Section \ref{sec:Con} provides the conclusion and discussion.

\section{The precessing magnetar central engine model}
\label{sec:precessing magnetar}
Most previous studies have assumed a fixed magnetic inclination angle for the magnetar central engine model \citep{Kasen&Bildsten2010,Woosley2010}. However, the newborn magnetar formed after the collapse of a massive star may deviate from spherical symmetry \citep{Suvorov&Kokkotas2020,Zou&Liang2022,ZhangBiao2024}. When the rotation axis of the deformed magnetar is misaligned with its principal axis, the magnetar undergoes precession, which causes the magnetic inclination angle to change over time \citep{Zanazzi&Lai2015,ZhangBiao2024}. In this work, we adopt a precessing deformed magnetar with a time-varying magnetic inclination angle as the central engine for SLSNe.

The rotational energy of the magnetar is given by
\begin{equation}
    E_{\rm rot} = \dfrac{1}{2} I \Omega^{2},
\label{eq:E_rot}
\end{equation}
where $I=(2/5)MR^{2}$ represents the moment of inertia of a magnetar with mass $M$ and radius $R$, and $\Omega$ is its angular frequency.

When the magnetar's rotational energy is primarily lost through magnetic dipole radiation, one has \citep{Suvorov&Kokkotas2020,Suvorov&Kokkotas2021} 
\begin{equation}
    -\dfrac{dE_{\rm rot}}{dt}=-I \Omega \dot{\Omega}=L_{\rm EM}=\dfrac{B_{\rm p}^{2}R^{6}\Omega^{4}}{6c^{3}}\lambda(\alpha),
\label{eq:dE_rot_dt}
\end{equation}
where $L_{\rm EM}$ is the magnetic dipole radiation luminosity, $B_{\rm p}$ is the surface magnetic field strength of the magnetar, and $c$ is speed of light.

The $\lambda$ in equation (\ref{eq:dE_rot_dt}) is the magnetosphere factor, which is a function of the magnetic inclination angle $\alpha$. The expression for the magnetosphere factor varies depending on the different magnetospheric environments of the magnetar \citep{Arzamasskiy2015}. When the magnetar's magnetosphere is a vacuum magnetosphere, the expression for the magnetosphere factor is $\lambda(\alpha)=\sin^{2}\alpha$ \citep{Ostriker&Gunn1969,Michel&Goldwire1970}. However, the magnetar is likely surrounded by a plasma-filled environment \citep{Goldreich&Julian1969}. When the magnetar's magnetosphere is plasma-filled, the expression for the magnetosphere factor is given by $\lambda(\alpha) \simeq 1 + \sin^2\alpha$ \citep{Spitkovsky2006,Kalapotharakos&Contopoulos2009,Arzamasskiy2015,Philippov2015}. In this work, we adopt the same magnetosphere factor expression as \cite{Suvorov&Kokkotas2020}, \cite{Zou&Liang2022}, and \cite{ZhangBiao2024}, which is 
\begin{equation}
    \lambda(\alpha)=1+\delta\sin^{2}\alpha,
\label{eq:lambda_alpha}
\end{equation}
where $\delta$ is a parameter related to the characteristics and physics of the magnetosphere, and it satisfies $\left| \delta \right| \leq 1$ \citep{Suvorov&Kokkotas2020,Zou&Liang2022,Arzamasskiy2015}.

\cite{Zanazzi&Lai2015}, \cite{GaoYong2023}, and \cite{ZhangBiao2024} provided the evolution of the magnetic inclination angle \( \alpha \) over time for a precessing deformed magnetar, which is expressed as
\begin{equation}
    \cos\alpha=\sin\chi\sin\theta\cos(\Omega_{\rm P}t)+\cos\chi\cos\theta,
\label{eq:cos_alpha}
\end{equation}
where $\chi$ is the angle between the magnetic axis and the principal axis corresponding to the largest moment of inertia, $\theta$ is the angle between the angular velocity vector and this principal axis, and $\Omega_{\rm P}$ is the precession frequency.

By solving the system of equations (\ref{eq:dE_rot_dt}), (\ref{eq:lambda_alpha}), and (\ref{eq:cos_alpha}), we can obtain the electromagnetic spin-down luminosity that incorporates the effects of magnetar precession, which is
\begin{eqnarray}
    L_{\rm sd,pre} &=& \dfrac{B_{\rm p}^{2}R^{6}\Omega_{0}^{4}}{6c^{3}} \times \left( 1 + \delta\sin^{2}\alpha \right) \nonumber \\ 
    &\times& \left\{ 1 + \dfrac{t \left[ 1 + \delta \left( 1 - \cos^{2}\chi \cos^{2}\theta - 0.5 \sin^{2}\chi \sin^{2}\theta \right) \right]}{3I c^{3} / (B_{\rm p}^{2} R^{6} \Omega_{0}^{2})} \right. \nonumber \\
    &-& \left. \dfrac{\delta\left[ 2 \cos\chi \cos\theta + 0.5 \sin\chi \sin\theta \cos(\Omega_{\rm P}t) \right]}{\{\Omega_{\rm P} / [\sin(\Omega_{\rm P}t)\sin\chi\sin\theta]\} \cdot [3I c^{3} / (B_{\rm p}^{2} R^{6} \Omega_{0}^{2})]} \right\}^{-2}
\label{eq:Lsd_pre}
\end{eqnarray}
where $\sin\alpha$ can be derived from equation (\ref{eq:cos_alpha}), $\Omega_{0}=2\pi/P_{0}$ is the initial angular frequency, and the precession frequency $\Omega_{P}$ is given by \citep{Suvorov&Kokkotas2020,Suvorov&Kokkotas2021} 
\begin{equation}
    \Omega_{\rm p}=\epsilon\cos\theta\Omega_{0}\left[1+\dfrac{t}{3I c^{3} / (B_{\rm p}^{2} R^{6} \Omega_{0}^{2})}\right]^{-1/2}.
\label{eq:Omega_p}
\end{equation}
Here, $\epsilon$ is the ellipticity of the magnetar. The precession-modified spin-down luminosity equation (\ref{eq:Lsd_pre}) that we derive here is consistent with equation (6) in \cite{Suvorov&Kokkotas2020}. 

\begin{figure*}
\centering
\includegraphics[width=0.48\textwidth, angle=0]{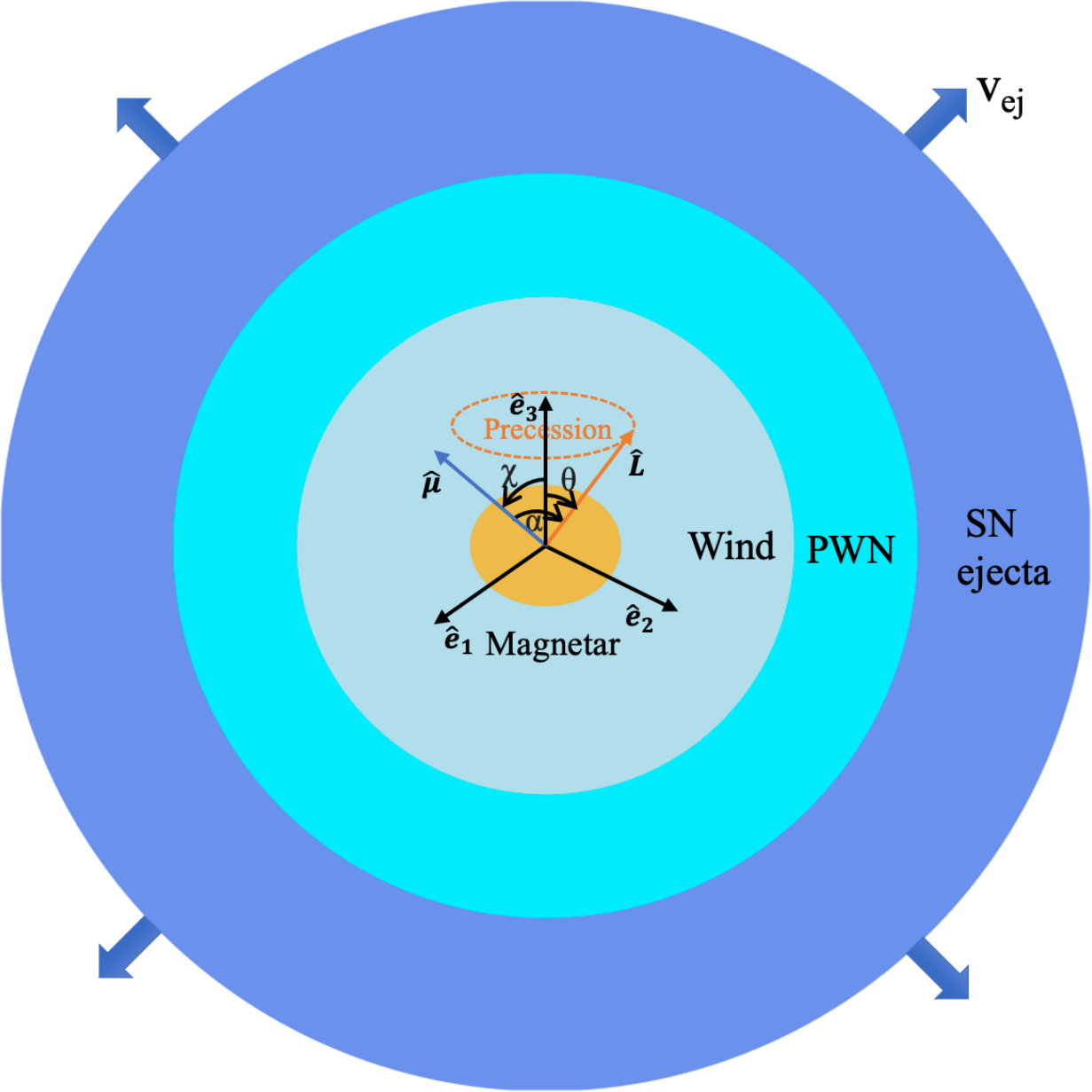}
\includegraphics[width=0.48\textwidth, angle=0]{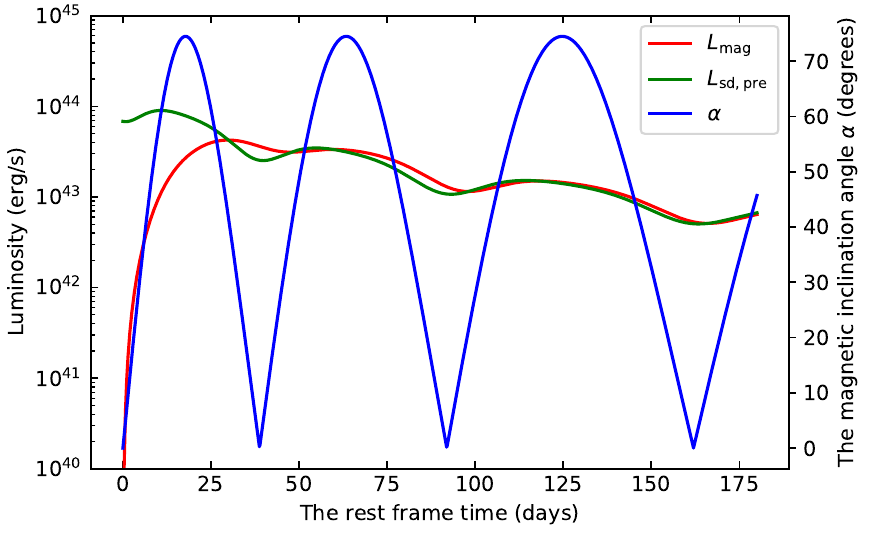}
\caption{The schematic diagram of our precessing magnetar-powered model. Left panel: Cartoon illustration of our precessing magnetar central engine model. A precessing, deformed magnetar is shown as the orange ellipse. The light blue, cyan, and royal blue regions represent the wind, the pulsar wind nebula (PWN), and the supernova ejecta, respectively. $\boldsymbol{\hat{e}_{1}}$, $\boldsymbol{\hat{e}_{2}}$, and $\boldsymbol{\hat{e}_{3}}$ are the coordinate axes in the corotating body frame (also the three principal axes of inertia). The unit dipole moment $\boldsymbol{\hat{\mu}}$ is fixed, with its polar angle $\chi$. The unit angular momentum $\boldsymbol{\hat{L}}$ precesses around $\boldsymbol{\hat{e}_{3}}$, with its polar angle $\theta$. $\alpha$ is the magnetic inclination angle. Right panel: The time evolution of the magnetic inclination angle $\alpha$ (blue line), precession-modified spin-down luminosity $L_{\rm sd,pre}$ (green line), and the corresponding bolometric luminosity of the SLSN $L_{\rm mag}$ (red line) for a set of typical parameters.}
\label{fig:The_model_schematic_diagram}
\end{figure*}

After taking into account the leakage effect of high-energy photons and the precession of the magnetar, the bolometric luminosity of an SLSN is given by \citep{Arnett1982,WangShan-Qin2015}
\begin{eqnarray} 
    L_{\rm mag}(t)=&e&^{-\left(\dfrac{t}{t_{\rm diff}}\right)^{2}} \left(1-e^{-At^{-2}}\right) \nonumber \\ 
    &\times& \int_{0}^{t} 2L_{\rm sd,pre}(t^{\prime})\dfrac{t^{\prime}}{t_{\rm diff}}e^{\left(\dfrac{t^{\prime}}{t_{\rm diff}}\right)^{2}}\dfrac{dt^{\prime}}{t_{\rm diff}}.
\label{eq:L_mag_t}
\end{eqnarray}
Here, we adopt a precessing deformed magnetar as the central engine for the SLSN, and thus, the energy injected into the supernova ejecta is given by the spin-down luminosity $L_{\rm sd,pre}$ of the precessing deformed magnetar. $t_{\rm diff}$ is the diffusion time, which can be expressed as
\begin{equation}
    t_{\rm diff}=\left(\dfrac{2\kappa M_{\rm ej}}{\beta c v_{\rm ej}}\right)^{1/2},
\label{eq:t_diff}
\end{equation}
where $M_{\rm ej}$ and $v_{\rm ej}$ are the ejecta mass and ejecta velocity, repectively, c denotes the speed of light, $\beta\simeq13.8$ is a constant, $\kappa$ represents the optical opacity, which is taken as ${\rm 0.1\,cm^2\,g^{-1}}$ in this work \citep{Inserra2013,YuYun-Wei2017}. The term $\left(1-e^{-At^{-2}}\right)$ represents the trapping factor for high-energy photons, where $A$ is given by \citep{WangShan-Qin2015}
\begin{equation}
    A=\dfrac{3\kappa_{\gamma}M_{\rm ej}}{4\pi v_{\rm ej}^{2}},
\label{eq:A}
\end{equation}
and $\kappa_{\gamma}$ is the gamma ray opacity. 

In many studies of non-precessing magnetar-powered models, such as those by \cite{Kasen&Bildsten2010}, \cite{Woosley2010}, \cite{Inserra2013}, and \cite{Metzger2015}, it is assumed that the magnetic dipole luminosity of the magnetar is fully thermalized to power the SLSNe. In contrast, other studies of non-precessing magnetar-powered models, such as those by \cite{WangShan-Qin2015}, \cite{Chen2015}, and \cite{Nicholl2017}, take into account that a fraction of high-energy photons can escape from the ejecta at late times. In these cases, the thermalization efficiency of the energy injected by the magnetar into the supernova ejecta is given by
\begin{equation}
    \eta_{\rm th}=1-e^{-At^{-2}}.
\label{eq:eta_th}
\end{equation}
Our precessing magnetar-powered model takes into account the late-time leakage of high-energy photons. Therefore, we adopt the same expression for the thermalization efficiency as in the studies mentioned above, namely equation (\ref{eq:eta_th}). The expression for thermalization efficiency adopted here is the same as that used in \cite{Nicholl2017}, \cite{Moriya2018}, and \cite{Vurm2021}. A thermalized luminosity can be defined as
\begin{equation}
    L_{\rm th}=L_{\rm sd,pre}\times\eta_{\rm th}=L_{\rm sd,pre}(1-e^{-At^{-2}}).
\label{eq:L_th}
\end{equation}
Both the non-precessing magnetar-powered model assuming $100\%$ thermalization efficiency adopted by \cite{Inserra2013}, and the models employing the thermalization efficiency given by equation (\ref{eq:eta_th}), such as those by \cite{WangShan-Qin2015} and \cite{Nicholl2017}, assume homologous expansion of the supernova ejecta with a constant expansion velocity. Our precessing magnetar-powered model adopts the same assumption, namely that the supernova ejecta expands homologously with a constant velocity.

To obtain the multi-band light curves of the SLSNe, we adopt the expressions for the photospheric temperature and radius from \cite{Nicholl2017}, which are given by:
\begin{equation}
\begin{scriptstyle}
T_{\rm pho}(t) = 
\begin{cases}
\left(\dfrac{L_{\rm mag}(t)}{4\pi \sigma v_{\rm ej}^{2}t^{2}}\right)^{1/4}, & \left(\dfrac{L_{\rm mag}(t)}{4\pi \sigma v_{\rm ej}^{2}t^{2}}\right)^{1/4} > T_{\rm f} \\
T_{\rm f}, & \left(\dfrac{L_{\rm mag}(t)}{4\pi \sigma v_{\rm ej}^{2}t^{2}}\right)^{1/4} \leq T_{\rm f}
\end{cases}
\label{eq:T_pho}
\end{scriptstyle}
\end{equation}
and
\begin{equation}
\begin{scriptstyle}
R_{\rm pho}(t) = 
\begin{cases}
v_{\rm ej}t, & \left(\dfrac{L_{\rm mag}(t)}{4\pi \sigma v_{\rm ej}^{2}t^{2}}\right)^{1/4} > T_{\rm f} \\
\left(\dfrac{L_{\rm mag}(t)}{4\pi \sigma T_{\rm f}^{4}}\right)^{1/2}, & \left(\dfrac{L_{\rm mag}(t)}{4\pi \sigma v_{\rm ej}^{2}t^{2}}\right)^{1/4} \leq T_{\rm f}
\end{cases}.
\label{eq:R_pho}
\end{scriptstyle}
\end{equation}
Here, $\sigma$ denotes the Stefan-Boltzmann constant, and $T_{\rm f}$ represents the final plateau temperature inspired by the observations \citep{Inserra2013,Nicholl2017}. By combining equations (\ref{eq:cos_alpha})-(\ref{eq:R_pho}), the bolometric luminosity and multi-band light curves of an SLSN powered by a precessing deformed magnetar can be obtained.

\begin{figure*}
\centering
\includegraphics[width=0.48\textwidth, angle=0]{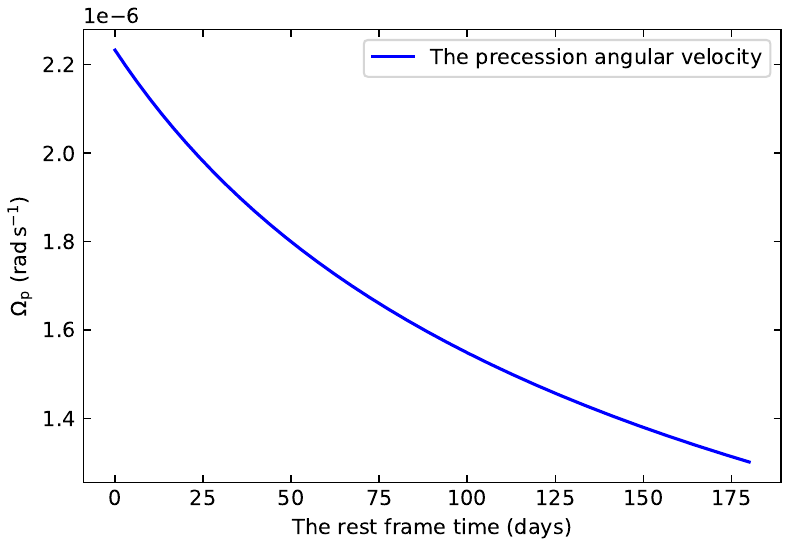}
\includegraphics[width=0.48\textwidth, angle=0]{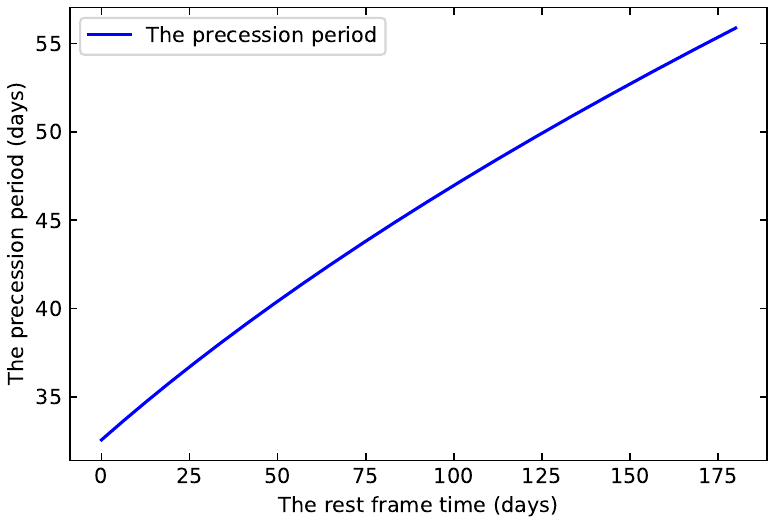}
\caption{The temporal evolution of the precession angular velocity (left panel) and precession period (right panel) of the magnetar, using the same set of parameters as in the right panel of Figure \ref{fig:The_model_schematic_diagram}. }
\label{fig:Precession_angular_velocity_period}
\end{figure*}

The radiation mechanism by which the magnetic dipole energy of the precessing magnetar is transferred to the ejecta in our model is the same as that used in non-precessing magnetar-powered models. The general process is as follows: the rotational energy lost by the magnetar initially manifests as a Poynting flux-dominated wind. Subsequently, magnetic reconnection and related processes within the Poynting flux-dominated wind convert the magnetic energy into the kinetic energy of particles, resulting in a relativistic wind composed of electron–positron pairs. When the magnetar wind catches up with and collides with the supernova ejecta, it drives a forward shock propagating outward into the ejecta and a reverse shock (commonly referred to as the termination shock) propagating inward into the wind. The shocked material between the forward and the termination shocks forms the pulsar wind nebula (PWN) \citep{Kotera2013}. Finally, photons from the PWN are absorbed and thermalized by the supernova ejecta \citep{Vurm2021,Li2024}. The wind is terminated at the radius where its ram pressure is balanced by the pressure of the PWN \citep{Gaensler&Slane2006,Yang&Dai2019}.

In the left panel of Figure \ref{fig:The_model_schematic_diagram}, we show a schematic illustration of our precessing magnetar central engine model. In the right panel of Figure \ref{fig:The_model_schematic_diagram}, we present the time evolution of the magnetic inclination angle $\alpha$, the precession-modified spin-down luminosity $L_{\rm sd,pre}$, and the bolometric luminosity of the SLSN $L_{\rm mag}$ for a set of typical parameters: $M_{\rm ej} = 6.03\,M_{\odot}$, $P_{\rm 0} = 6.34\,{\rm ms}$, $B_{\rm p} = 1.07 \times 10^{14}\,{\rm G}$, $v_{\rm ej} = 0.94 \times 10^{9}\,{\rm cm\,s^{-1}}$, ${\rm log_{10}(\kappa_{\gamma}/cm^2g^{-1})} = 0.44$, $T_{\rm f} = 5184.57\,{\rm K}$, $\epsilon = 2.83 \times 10^{-9}$, $\theta = 0.65$, $\chi = 0.65$, and $\delta = 1.00$. Our precessing magnetar-powered model explains the late-time bumps in the light curves of SLSNe-I without invoking any additional energy beyond the magnetic dipole radiation of the precessing magnetar. The magnetic dipole luminosity generated by a magnetar depends on the magnetic inclination angle $\alpha$. In non-precessing magnetar-powered models (i.e., original magnetar-powered models), $\alpha$ is typically assumed to be a fixed value. The magnetic dipole luminosity from a non-precessing magnetar decreases continuously and smoothly over time. As a result, the energy injection rate into the supernova ejecta from a non-precessing magnetar also decreases steadily. From equation (\ref{eq:Lsd_pre}), it can be seen that the magnetic dipole luminosity of a precessing magnetar is dependent on the magnetic inclination angle $\alpha$. During the precession of the magnetar, the inclination angle $\alpha$ undergoes quasi-periodic variations (i.e., it increases and decreases quasi-periodically), as shown by the blue line in the right panel of Figure \ref{fig:The_model_schematic_diagram}. According to equation (\ref{eq:Lsd_pre}), when $\alpha$ increases, the magnetic dipole luminosity of the precessing magnetar increases accordingly; conversely, when $\alpha$ decreases, the luminosity decreases. This co-evolution of the magnetic inclination angle and magnetic dipole luminosity of the precessing magnetar can be seen from the simultaneous increase and decrease of the blue line (representing the magnetic inclination angle $\alpha$) and the green line (representing the magnetic dipole luminosity) in the right panel of Figure \ref{fig:The_model_schematic_diagram}. Therefore, unlike the continuously and smoothly decreasing magnetic dipole luminosity produced by a non-precessing magnetar, the magnetic dipole luminosity produced by a precessing magnetar exhibits quasi-periodic increases and decreases superimposed on an overall declining trend, as shown by the green line in the right panel of Figure \ref{fig:The_model_schematic_diagram}. In our work, the energy injection rate for the ejecta is equal to the magnetic dipole luminosity produced by the precessing magnetar. Therefore, the quasi-periodic increase and decrease of the magnetic dipole luminosity produced by the precessing magnetar result in a corresponding quasi-periodic increase and decrease in the energy injection rate into the supernova ejecta. In our precessing magnetar-powered model, the appearance of bumps in the light curve is attributed to the quasi-periodic increase and decrease of the energy injection rate into the supernova ejecta. An increase in the energy injection rate corresponds to the rising phase of a bump, while a decrease in the energy injection rate corresponds to the declining phase of a bump. This correlated behavior between the energy injection rate and the SLSN light curve can be clearly seen from the green line (representing the energy injection rate) and the red line (representing the SLSN light curve) in the right panel of Figure \ref{fig:The_model_schematic_diagram}. Equation (\ref{eq:Omega_p}) provides the expression for the precession angular velocity in our precessing magnetar-powered model. Based on this equation and using the same set of parameters as in the right panel of Figure \ref{fig:The_model_schematic_diagram}, we present the temporal evolution of the magnetar's precession angular velocity (left panel) and precession period (right panel) in Figure \ref{fig:Precession_angular_velocity_period}. It can be seen that the precession period of the magnetar increases continuously over time.

The impact of our precessing magnetar central engine model on the SLSNe light curves is primarily manifested in the late phases, for the following reasons. On the one hand, in the early stages, the diffusion time of photons in the supernova ejecta is relatively long, which causes the variations in the energy injection rate due to magnetar precession to be significantly or even completely smoothed out by the diffusion process. As time passes, the diffusion timescale of photons in the supernova ejecta decreases. Therefore, the smoothing effect of the diffusion process on the variations in the energy injection rate caused by magnetar precession becomes weaker. As a result, in the late stages, the variations in the energy injection rate due to magnetar precession will manifest in the light curves of SLSNe. On the other hand, the precession period of the magnetar continues to increase, which in turn causes the timescale of the variations in the energy injection rate due to magnetar precession to also increase. This further reduces the impact of the smoothing effect of the diffusion process on these variations in the late stages. Therefore, the effects of magnetar precession are primarily manifested in the late stages of the SLSNe light curves. As seen in the right panel of Figure \ref{fig:The_model_schematic_diagram}, in the early stages, the spin-down luminosity (which is also the energy injection rate) produced by the precessing magnetar varies. However, due to the smoothing effect of photon diffusion, the bolometric light curve of the SLSN does not reflect the variations caused by precession. In the late stages, the influence of the photon diffusion smoothing effect is significantly reduced, and the variations in the spin-down luminosity produced by the precessing magnetar are reflected in the bolometric light curve of the SLSN.

\section{Sample selection and light curve fitting}
\label{sec:Lightcurv}

\begin{deluxetable}{lcl}
\label{tab:basic_information}
\tablecaption{Basic information for the 6 SLSNe-I with post-peak bumps in our sample.}
\tablehead{
\multicolumn{1}{l}{Name} &
\colhead{\centering Redshift} &
\multicolumn{1}{l}{Photometry Reference} 
}
\startdata
\object{SN 2018kyt} & $0.1080$ & \cite{Chen2023_photometry} \\
\object{SN 2019lsq} & $0.1295$ & \cite{Chen2023_photometry} \\ 
\object{SN 2019hge} & $0.0866$ & \cite{Chen2023_photometry}; \\ & \, & Bright Transient Survey explorer \\
\object{SN 2019stc} & $0.1178$ & \cite{Gomez2021}; \cite{DongXiao-Fei2023}; \\ & \, & \cite{ZhuJin-Ping2024} \\
\object{SN 2021mkr} & $0.28$ & \cite{DongXiao-Fei2023}; \\ & \, & Bright Transient Survey explorer \\
\object{PS1-12cil}  & $0.32$ & \cite{Lunnan2018} \\
\enddata
\end{deluxetable}

\setlength{\tabcolsep}{1.2pt}
\begin{deluxetable*}{lccccccccccc}
\label{tab:emcee_parameters}
\tablecaption{The best-fit model parameters and priors of the free parameters of the model.}
\tablehead{
\multicolumn{1}{l}{Name} &
\colhead{\centering $M_{\rm ej}$} &
\colhead{\centering $P_{\rm 0}$} &
\colhead{\centering $B_{\rm p}$} &
\colhead{\centering $v_{\rm ej}$} &
\colhead{\centering ${\rm log_{10}(\kappa_{\gamma}/cm^2g^{-1})}$} &
\colhead{\centering $T_{\rm f}$} &
\colhead{\centering $t_{\rm shift}$} &
\colhead{\centering $\epsilon$} &
\colhead{\centering $\theta$} &
\colhead{\centering $\chi$} &
\colhead{\centering $\delta$} 
\\
\multicolumn{1}{l}{} &
\colhead{\centering ($M_{\odot}$)} &
\colhead{\centering (${\rm ms}$)} &
\colhead{\centering (${\rm 10^{14}\,G}$)} &
\colhead{\centering (${\rm 10^{9}\,cm\,s^{-1}}$)} &
\colhead{\centering \,} &
\colhead{\centering (${\rm K}$)} &
\colhead{\centering (${\rm days}$)} &
\colhead{\centering (${\rm \times10^{-9}}$)} &
\colhead{\centering (${\rm rad}$)} &
\colhead{\centering (${\rm rad}$)} &
\colhead{\centering } 
}
\startdata
\object{SN 2018kyt}  & $3.08^{+0.29}_{-0.27}$ & $5.62^{+0.06}_{-0.06}$ & $1.12^{+0.04}_{-0.04}$ & $0.59^{+0.02}_{-0.02}$ & $-1.30^{+0.06}_{-0.06}$ & $7441.58^{+40.58}_{-43.11}$ & $6.21^{+0.54}_{-0.47}$ & $2.36^{+0.22}_{-0.19}$ & $0.98^{+0.06}_{-0.06}$ & $0.96^{+0.06}_{-0.06}$ & $0.79^{+0.06}_{-0.05}$ \\
\object{SN 2019lsq} &  $5.02^{+0.12}_{-0.12}$ & $3.46^{+0.02}_{-0.02}$ & $1.15^{+0.01}_{-0.01}$ & $0.76^{+0.01}_{-0.01}$ & $0.54^{+0.80}_{-0.79}$ & $9217.46^{+84.69}_{-74.93}$ & $4.09^{+0.11}_{-0.11}$ & $2.80^{+0.23}_{-0.20}$ & $1.04^{+0.04}_{-0.04}$ & $1.04^{+0.05}_{-0.05}$ & $0.97^{+0.02}_{-0.03}$ \\ 
\object{SN 2019hge} &  $4.86^{+0.13}_{-0.12}$ & $5.43^{+0.01}_{-0.01}$ & $1.01^{+0.04}_{-0.03}$ & $0.45^{+0.00}_{-0.00}$ & $-1.99^{+0.01}_{-0.00}$ & $5584.65^{+73.24}_{-62.09}$ & $11.87^{+0.18}_{-0.18}$ & $5.22^{+0.19}_{-0.18}$ & $0.88^{+0.04}_{-0.04}$ & $0.87^{+0.04}_{-0.04}$ & $0.99^{+0.01}_{-0.01}$ \\
\object{SN 2019stc} &  $6.03^{+0.29}_{-0.27}$ & $6.34^{+0.02}_{-0.02}$ & $1.07^{+0.01}_{-0.01}$ & $0.94^{+0.02}_{-0.02}$ & $0.44^{+0.82}_{-0.81}$ & $5184.57^{+25.93}_{-26.04}$ & $7.48^{+0.27}_{-0.25}$ & $2.83^{+0.04}_{-0.04}$ & $0.65^{+0.02}_{-0.02}$ & $0.65^{+0.02}_{-0.02}$ & $1.00^{+0.00}_{-0.00}$ \\
\object{SN 2021mkr} &  $18.54^{+4.22}_{-3.86}$ & $1.97^{+0.07}_{-0.09}$ & $0.24^{+0.04}_{-0.03}$ & $1.42^{+0.14}_{-0.12}$ & $0.43^{+0.89}_{-0.92}$ & $9004.40^{+370.64}_{-322.70}$ & $17.44^{+2.09}_{-2.27}$ & $1.09^{+0.25}_{-0.17}$ & $0.72^{+0.22}_{-0.17}$ & $0.72^{+0.20}_{-0.18}$ & $0.80^{+0.12}_{-0.18}$ \\
\object{PS1-12cil} &  $2.07^{+0.05}_{-0.05}$ & $4.45^{+0.02}_{-0.02}$ & $1.48^{+0.01}_{-0.01}$ & $0.92^{+0.01}_{-0.01}$ & $-0.97^{+0.02}_{-0.02}$ & $7339.52^{+47.34}_{-46.68}$ & $5.08^{+0.15}_{-0.14}$ & $4.01^{+0.04}_{-0.04}$ & $0.49^{+0.02}_{-0.02}$ & $0.49^{+0.02}_{-0.02}$ & $0.98^{+0.01}_{-0.02}$ \\
\hline
\object{Prior} & [0.1, 50] & [0.7, 20] & [0.01, 20] & [0.1, 4] & [-2, 2] & [1000, 20000] & [0, 25] & [0.01, 10000] & [0, $\pi/2$] & [0, $\pi/2$] & [-1, 1]
\enddata
\end{deluxetable*}

\cite{Hosseinzadeh2022} found 15 SLSNe-I with definite post-peak bumps and 11 SLSNe-I with possible post-peak bumps in a sample of 34 SLSNe-I. In addition, \cite{Chen2023} found 13 SLSNe-I with strong undulations and 4 SLSNe-I with weak undulations among the 73 SLSNe-I observed in the Phase-I Survey of Zwicky Transient Facility (ZTF). These results suggest that bumps are prevalent in SLSNe-I and play an important role in the study of these objects. Previous studies have primarily focused on investigating individual SLSNe-I with post-peak bumps using specific models (e.g., \citealt{Fiore2021,Gomez2021,West2023}). Recently, a few works have begun to employ the same model to study multiple SLSNe-I with post-peak bumps simultaneously \citep{DongXiao-Fei2023,ZhuJin-Ping2024}. However, such studies are still rare, and our work represents a new attempt in this area. In this Section, we fit the multi-band light curves of several SLSNe-I with post-peak bumps using the precessing deformed magnetar-powered SLSN model introduced in Section \ref{sec:precessing magnetar}. 

We searched the literature for SLSNe with the following characteristics: First, they can be classified as hydrogen-poor SLSNe-I based on their spectra. Second, they have well-sampled observational data in at least two photometric bands. Third, their light curves exhibit a primary peak, with no less than five data points in both the rising and declining phases of the peak. Fourth, their light curves exhibit one or more bumps that deviate from the smooth decline trend after the primary peak, with the rising and declining phases of the bumps forming complete structures. The sample we studied includes 6 SLSNe that meet these characteristics. It should be noted that only a subset of these SLSNe were selected as example cases for this study, and there may be additional SLSNe that also meet these characteristics but are not included in our sample. Table \ref{tab:basic_information} presents the basic information for the 6 SLSNe-I with post-peak bumps studied in this work. 
The photometry data for SN 2021mkr and partial photometry data for SN 2019hge are obtained from the Bright Transient Survey Explorer \citep{Fremling2020,Perley2020} catalog \footnote{\url{https://sites.astro.caltech.edu/ztf/bts/explorer.php}}.
SN 2018kyt, SN 2019lsq, SN 2019stc, SN 2021mkr, and PS1-12cil constitute the sample in \cite{DongXiao-Fei2023}. SN 2018kyt, SN 2019hge, and PS1-12cil are objects with definite bumps in the work of \cite{Hosseinzadeh2022}, while SN 2019lsq is an object with possible bumps in the same study. SN 2018kyt, SN 2019hge, SN 2019lsq, and SN 2019stc are objects with strong undulations in the work of \cite{Chen2023}. Therefore, the 6 SLSNe-I with post-peak bumps selected in our study represent a set of typical examples, which can be used to test whether our precessing deformed magnetar-powered model can explain SLSNe-I with post-peak bumps.

\begin{figure*}
\centering
\includegraphics[width=0.329\textwidth, angle=0]{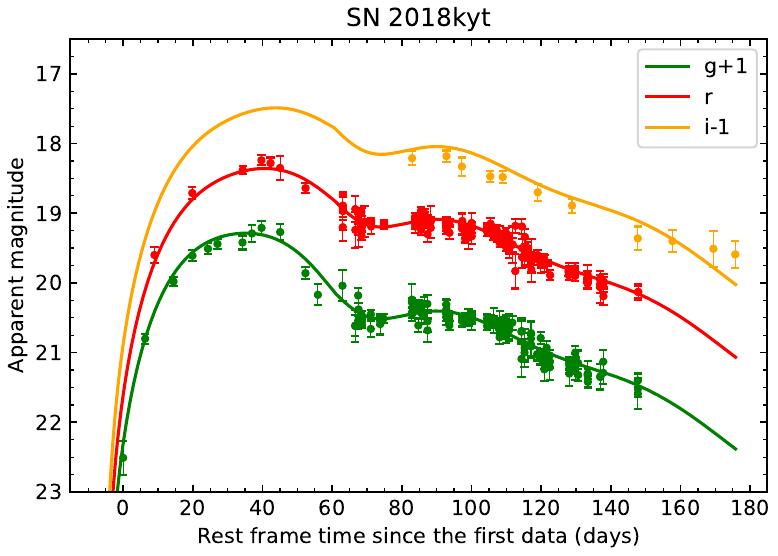}
\includegraphics[width=0.329\textwidth, angle=0]{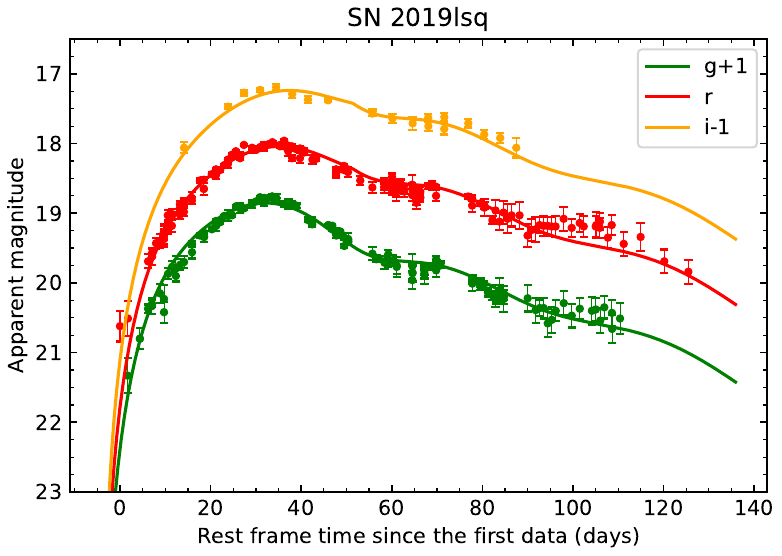}
\includegraphics[width=0.329\textwidth, angle=0]{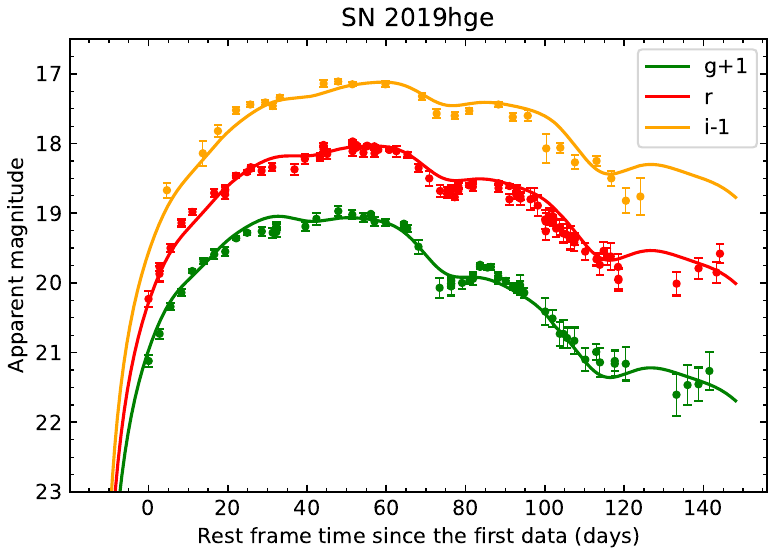}
\includegraphics[width=0.329\textwidth, angle=0]{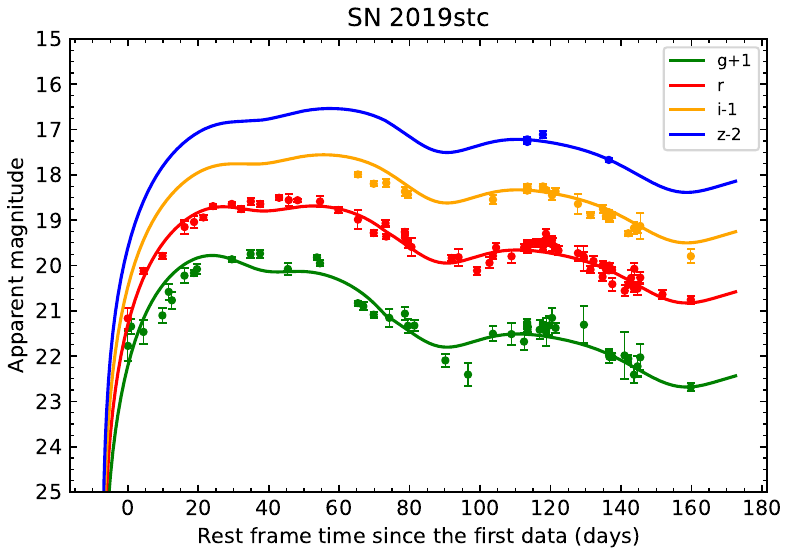}
\includegraphics[width=0.329\textwidth, angle=0]{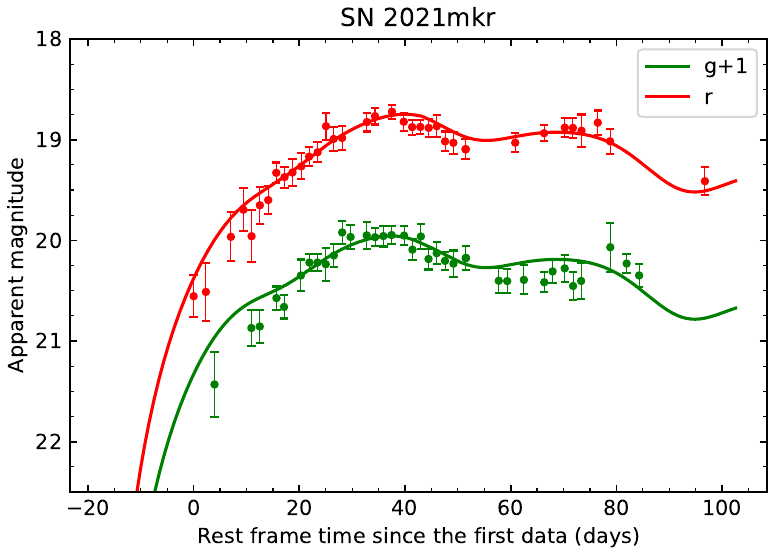}
\includegraphics[width=0.329\textwidth, angle=0]{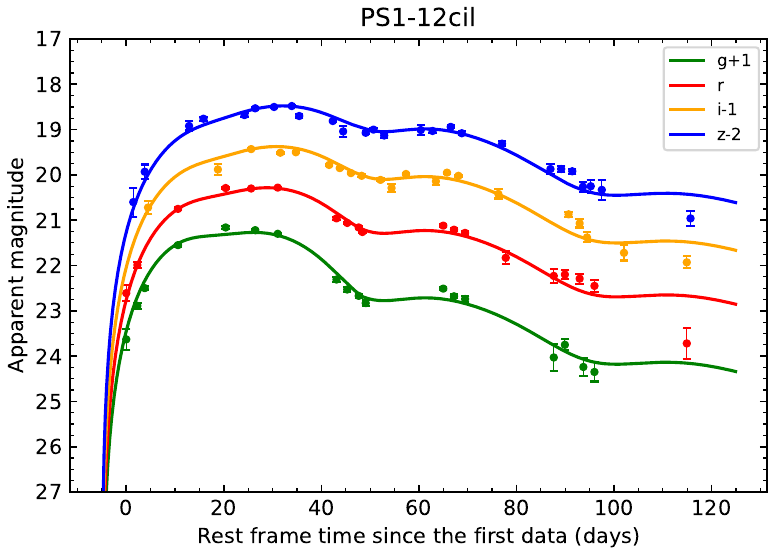}
\caption{Multi-band observational data for the 6 SLSNe-I with post-peak bumps in our sample, along with best-fitting curves from our precessing magnetar-powered model. Different bands are represented by different colors.}
\label{fig:emcee_lightcurve}
\end{figure*}

Taking the first observed data point as the zero-point, we introduce $t_{\rm shift}$ to shift the SLSN light curve leftward. The photometry data have been corrected for Galactic extinction in this work. The free parameters of our model include the initial spin period of the magnetar $P_0$, the surface magnetic field strength $B_{\rm p}$, the ejecta mass $M_{\rm ej}$, the ejecta velocity $v_{\rm ej}$, the opacity of gamma photons $\kappa_{\gamma}$, the final plateau temperature of the photosphere $T_{\rm f}$, the ellipticity of the magnetar $\epsilon$, the angle $\theta$ between the angular velocity vector and the principal axis of the maximum moment of inertia, the angle $\chi$ between the magnetic axis and the same principal axis, the parameter $\delta$ in the expression for the magnetospheric factor, and the time shift $t_{\rm shift}$ of the light curve. Table \ref{tab:emcee_parameters} presents the prior ranges for these model parameters. The priors for $M_{\rm ej}$, $P_0$, $B_{\rm p}$, $v_{\rm ej}$, $\kappa_{\gamma}$, $T_{\rm f}$, and $t_{\rm shift}$ are referenced from \cite{DongXiao-Fei2023} and \cite{ZhuJin-Ping2024}. The prior for $\epsilon$ is referenced from \cite{Zanazzi&Lai2020} and \cite{Zou&Liang2022}. The priors for $\chi$ and $\theta$ are referenced from \cite{Arzamasskiy2015}. The prior for $\delta$ is referenced from \cite{Suvorov&Kokkotas2020} and \cite{Zou&Liang2022}. To fit the multi-band photometry data with our model and constrain the free parameters of the model, we employed the Markov Chain Monte Carlo (MCMC) technique via the EMCEE Python package \citep{Foreman-Mackey2013}. Figure \ref{fig:emcee_lightcurve} shows the fitted light curves from our model to the multi-band observational data of the 6 SLSNe-I with post-peak bumps in our sample. Table \ref{tab:emcee_parameters} presents the best-fit model parameters obtained from the multi-band light curve fitting.

\section{Results and Discussion}
\label{sec:resulst}

\begin{figure*}
\centering
\includegraphics[width=0.48\textwidth, angle=0]{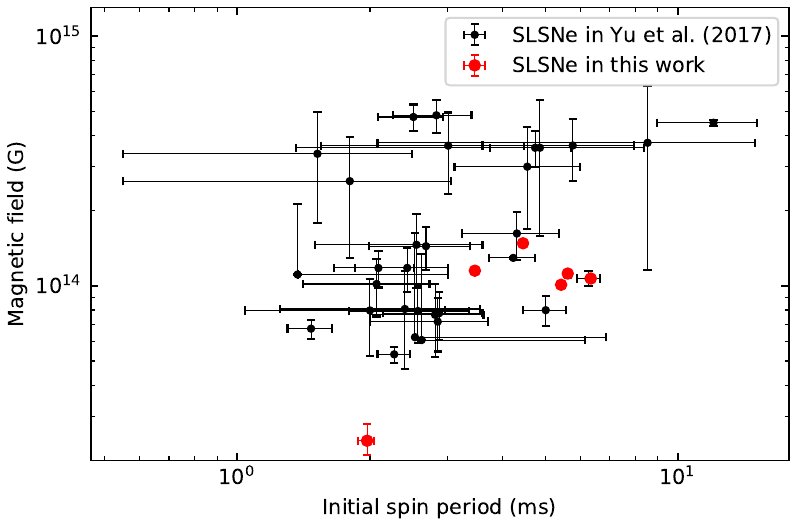}
\includegraphics[width=0.48\textwidth, angle=0]{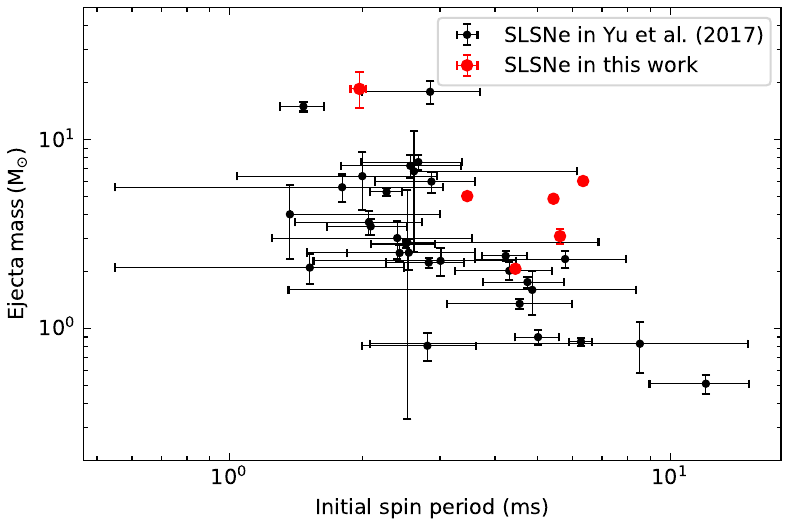}
\caption{Comparison of the parameters of the 6 SLSNe-I with post-peak bumps in our sample and the 31 SLSNe-I from \cite{YuYun-Wei2017}. Left panel: the $B_{\rm p}$ versus $P_0$ plane. Right panel: the $M_{\rm ej}$ versus $P_0$ plane. The red points represent the parameters of the SLSNe-I in this work, while the black points are the parameters of the SLSNe-I from \cite{YuYun-Wei2017}.}
\label{fig:Our_VS_Yu2017}
\end{figure*}

\begin{figure*}
\centering
\includegraphics[width=0.48\textwidth, angle=0]{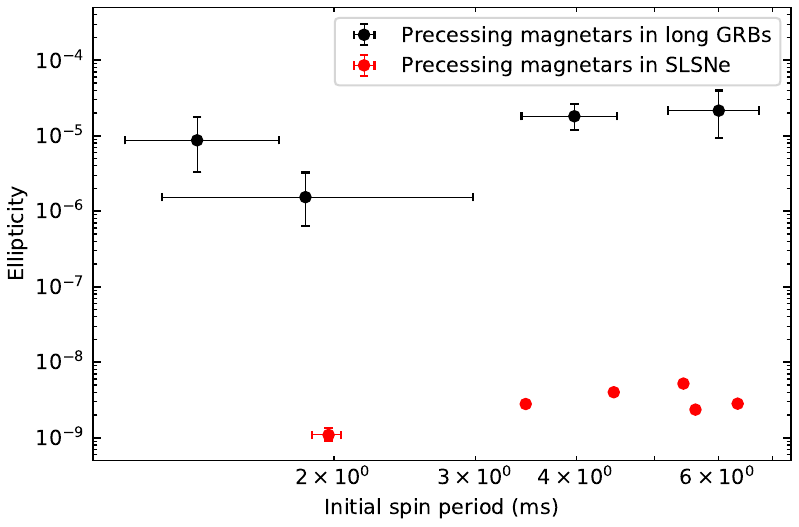}
\includegraphics[width=0.48\textwidth, angle=0]{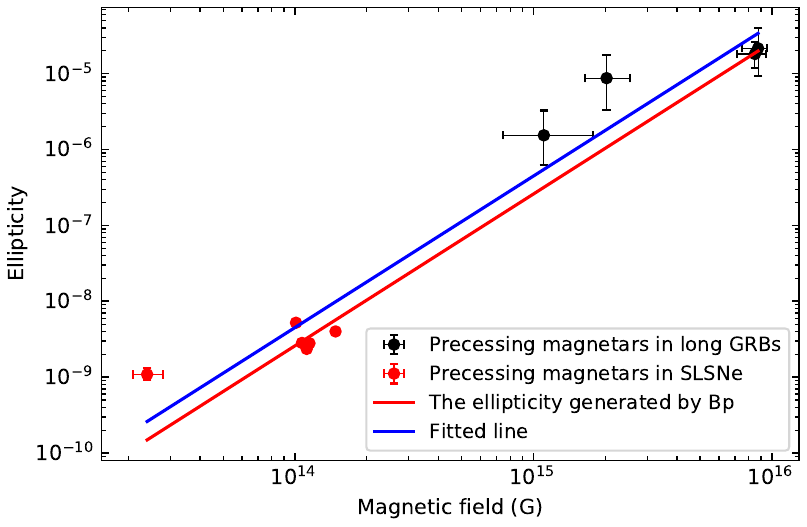}
\caption{Comparison of the parameters of the precessing magnetars in the SLSNe-I from this work and those in the long GRBs from \cite{ZhangBiao2024}. Left panel: the $\epsilon$ versus $P_0$ plane. Right panel: the $\epsilon$ versus $B_{\rm p}$ plane, where the solid blue line represents the fitted line to the red and blue data points. The red solid line represents the ellipticity generated by the dipole magnetic field $B_{\rm p}$ as given by equation (\ref{eq:epsilon_p}). The red points represent the parameters of the precessing magnetars in the SLSNe-I from this work, while the black points represent the parameters of the precessing magnetars in the long GRBs from \cite{ZhangBiao2024}.}
\label{fig:SLSNe_VS_GRBs}
\end{figure*}

\begin{figure*}
\centering
\includegraphics[width=0.48\textwidth, angle=0]{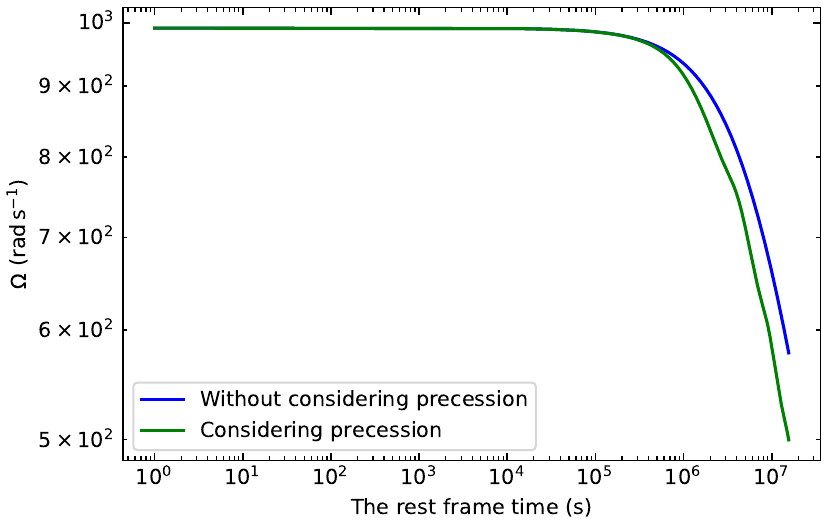}
\includegraphics[width=0.48\textwidth, angle=0]{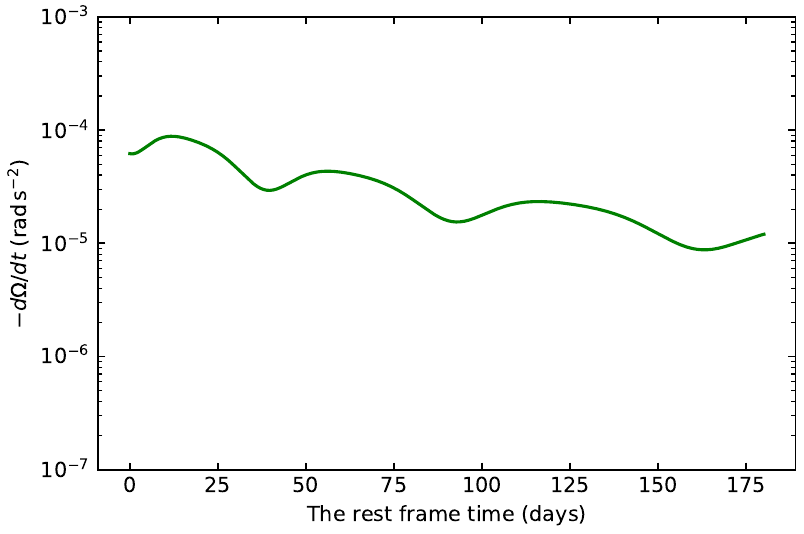}
\caption{Left panel: The evolution of the spin frequency $\Omega$ of the magnetar, with (green line) and without (blue line) considering the effects of precession, using the same set of typical parameters as in the right panel of Figure \ref{fig:The_model_schematic_diagram}. Right panel: The time evolution of $-d\Omega/dt$ for the precessing magnetar, using the same set of parameters as in the left panel.  }
\label{fig:The_spindown_timescale_rate}
\end{figure*}

\begin{figure*}
\centering
\includegraphics[width=0.48\textwidth, angle=0]{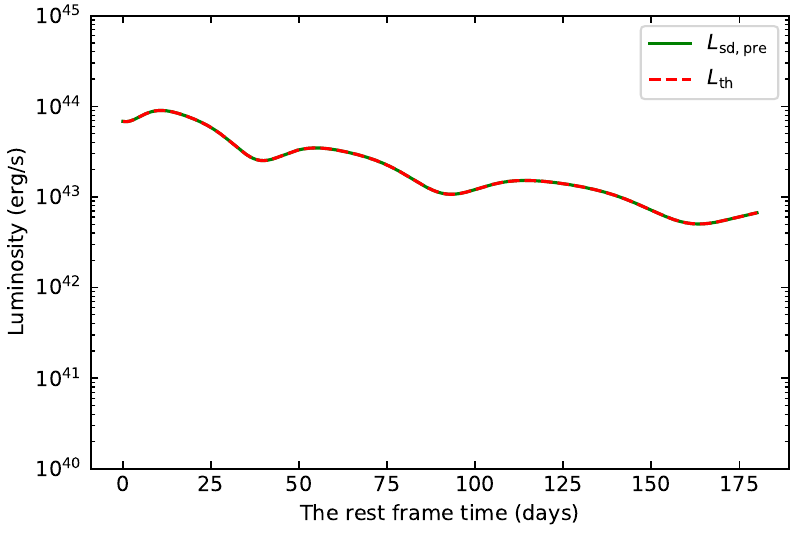}
\includegraphics[width=0.48\textwidth, angle=0]{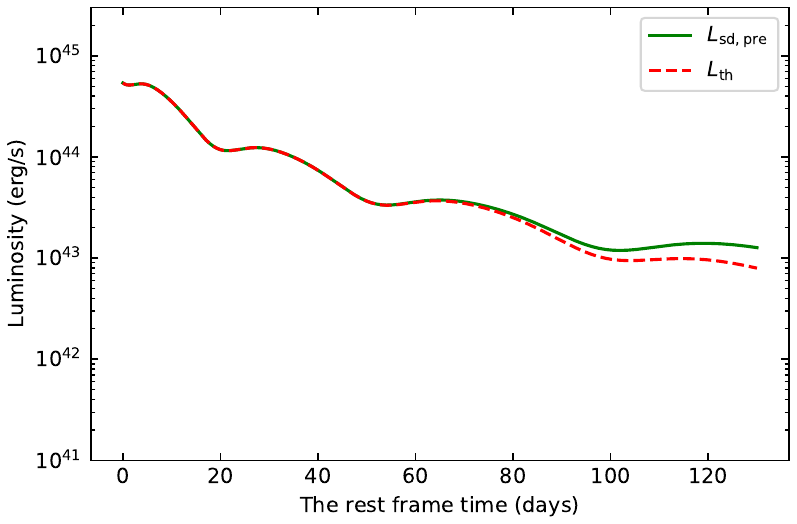}
\caption{Left panel: The magnetic dipole luminosity of the precessing magnetar $L_{\rm sd,pre}$ (green solid line) and the corresponding thermalized luminosity $L_{\rm th}$ (red dashed line) for SN 2019stc. The model parameter values are taken from Table \ref{tab:emcee_parameters}. Right panel: Same as the left panel, but for PS1-12cil. }
\label{fig:Thermal_luminosity}
\end{figure*}

As shown in Figure \ref{fig:emcee_lightcurve}, the theoretical light curves produced by our precessing magnetar-powered model well reproduce the observed light curves of the 6 SLSNe-I with post-peak bumps in our sample. It is worth noting that our precessing magnetar-powered model not only reproduces the observed light curves of SLSNe-I with a single post-peak bump, but also accounts for the observed light curves of SLSNe-I with multiple post-peak bumps. In contrast, some models proposed in previous studies, such as the circular orbit magnetar-star binary engine model \citep{ZhuJin-Ping2024} and the magnetar magnetic field enhancement model \citep{Chugai&Utrobin2022}, can only explain the observed light curves of SLSNe-I with a single post-peak bump. Our results suggest that the post-peak bumps in SLSNe-I may be caused by changes in the energy injected by the magnetar into the supernova ejecta due to the precession of the magnetar. Therefore, it may not be necessary to introduce additional energy sources to explain the post-peak bumps. Additionally, as shown in Figure \ref{fig:emcee_lightcurve}, the post-peak bumps in each SLSN-I of our sample appear almost simultaneously across different photometric bands, indicating that these post-peak bumps are wavelength-independent. 

Figure \ref{fig:Our_VS_Yu2017} presents the distribution of the parameters of the 6 SLSNe-I with post-peak bumps in our sample, together with the distribution of the parameters of 31 SLSNe-I from \cite{YuYun-Wei2017}. The left panel of Figure \ref{fig:Our_VS_Yu2017} shows the distribution of the dipole magnetic field strength $B_{\rm p}$ versus the initial spin period $P_0$, while the right panel of Figure \ref{fig:Our_VS_Yu2017} shows the distribution of the ejecta mass $M_{\rm ej}$ versus the initial spin period $P_0$. As shown in Figure \ref{fig:Our_VS_Yu2017}, the parameters of the 6 SLSNe-I with post-peak bumps in our sample are mixed with those of the 31 SLSNe-I from \cite{YuYun-Wei2017}, indicating that the parameters of the SLSNe-I with post-peak bumps in our sample fall within the typical parameter range for SLSNe-I. This is consistent with the findings of \cite{Hosseinzadeh2022} and \cite{DongXiao-Fei2023}, who concluded that there is no significant difference between the parameters of SLSNe-I with and without post-peak bumps.

\cite{ZhangBiao2024} interpreted the QPO signals observed in the X-ray afterglows of some long GRBs using a precessing magnetar as the central engine. Here, we compare the parameters of the precessing magnetars in the SLSNe-I from this work with those of the precessing magnetars in the long GRBs from \cite{ZhangBiao2024}. Figure \ref{fig:SLSNe_VS_GRBs} presents the parameter distributions of the precessing magnetars in the SLSNe-I from this work, together with those of the precessing magnetars in the long GRBs from \cite{ZhangBiao2024}. The left panel of Figure \ref{fig:SLSNe_VS_GRBs} shows the distribution of the ellipticity $\epsilon$ versus the initial spin period $P_0$ for the precessing magnetars. One can find that the ellipticity of the precessing magnetars in long GRBs is larger than that in SLSNe-I, and that the ellipticity of precessing magnetars in both long GRBs and SLSNe-I appears to be independent of the initial spin period $P_0$. One possible reason why the ellipticity appears to be independent of $P_0$ is that, as noted by \cite{Suvorov&Kokkotas2021}, the ellipticity considered here is the rotationally misaligned ellipticity (see equation 5 in \citealt{Suvorov&Kokkotas2021}), which only accounts for distortions that are misaligned with the rotation axis. It should be noted that the current sample size is limited, and these findings require verification with a larger sample in future studies. The right panel of Figure \ref{fig:SLSNe_VS_GRBs} shows the distribution of the ellipticity $\epsilon$ versus the dipole magnetic field strength $B_{\rm p}$ for the precessing magnetars. We find that the ellipticity $\epsilon$ and the dipole magnetic field strength $B_{\rm p}$, considering both the precessing magnetars in long GRBs and SLSNe-I, follow a relation expressed as:
\begin{equation}
    \log_{10}\epsilon=2.00\log_{10} B_{\rm p} - 36.29
\label{eq:log_epsilon_Bp}
\end{equation}
Equation (\ref{eq:log_epsilon_Bp}) is shown as the solid blue line in the right panel of Figure \ref{fig:SLSNe_VS_GRBs}. Furthermore, a Pearson correlation analysis of the parameters for the precessing magnetars in long GRBs and SLSNe-I yields a correlation coefficient ${\rm r = 0.97}$ and a p-value of $2.27 \times 10^{-6}$. This indicates that a stronger dipole magnetic field corresponds to a larger ellipticity. \cite{Zanazzi&Lai2015}, \cite{Zanazzi&Lai2020}, and \cite{WeiYu-Jia2022} have provided the ellipticity $\epsilon_{\rm p}$ caused by the dipole magnetic field $B_{\rm p}$ of the magnetar, which is given by
\begin{eqnarray}
    \epsilon_{\rm p}&=&\dfrac{3B^{2}_{\rm p}R^{5}}{20Ic^{2}} \nonumber \\
    &\approx& 2.592\times10^{-7}B^{2}_{\rm p,15}M^{-1}_{1.4}\left(\dfrac{R}{1.2\times10^{6}\,{\rm cm}}\right)^{3},
\label{eq:epsilon_p}
\end{eqnarray}
where $B_{\rm p,15}=B_{\rm p}/(10^{15}\,{\rm G})$ and $M_{1.4}=M/(1.4\,M_{\odot})$. Equation (\ref{eq:epsilon_p}) is shown as the red solid line in the right panel of Figure \ref{fig:SLSNe_VS_GRBs}. One can observe that the relationship between the ellipticity $\epsilon_{\rm p}$ and the dipole magnetic field $B_{\rm p}$, as given by equation (\ref{eq:epsilon_p}), can explain the distribution of ellipticity and dipole magnetic field in the right panel of Figure \ref{fig:SLSNe_VS_GRBs}. This suggests that the ellipticity of precessing magnetars in SLSNe-I from our study, as well as in long GRBs from \cite{ZhangBiao2024}, is generated by the magnetar's dipole magnetic field. The fact that both SLSNe-I and long GRBs satisfy the relation described by Equation (\ref{eq:log_epsilon_Bp}) may serve as evidence that they share a common central engine. These findings also need to be confirmed using larger sample sizes in future work.

We use the same set of typical parameters as in the right panel of Figure \ref{fig:The_model_schematic_diagram}, and in the left panel of Figure \ref{fig:The_spindown_timescale_rate}, we show the evolution of the spin frequency $\Omega$ for the magnetar, both without considering the precession effect (blue line) and with the precession effect taken into account (green line). As shown in the left panel of Figure \ref{fig:The_spindown_timescale_rate}, the end times of the plateau for the green and blue lines are nearly the same. However, after the plateau, the decay rate of the green line is faster than that of the blue line. Additionally, after the plateau, the green line exhibits bumpy features superimposed on the overall declining trend. This indicates that precession causes the magnetar's rotational frequency $\Omega$ to decay more rapidly after the plateau and leads to the appearance of bumpy features. In the right panel of Figure \ref{fig:The_spindown_timescale_rate}, we show the time evolution of $-d\Omega/dt$ for the precessing magnetar under the same set of parameters. It can be found that the precession leads to quasi-periodic variations in the spin-down rate. However, the period of precession increases rapidly, and the resulting observable consequences remain to be explored in future studies.

Based on equations (\ref{eq:Lsd_pre}) and (\ref{eq:L_th}), and adopting the model parameter values listed in Table \ref{tab:emcee_parameters}, we present the magnetic dipole luminosity of the precessing magnetar and the corresponding thermalized luminosity for SN 2019stc (left panel) and PS1-12cil (right panel) in Figure \ref{fig:Thermal_luminosity}. It is found that the thermalized luminosity of SN 2019stc remains equal to the magnetic dipole luminosity throughout the entire observational period. The cases of SN 2019lsq and SN 2021mkr are similar to that of SN 2019stc. For PS1-12cil, the thermalized luminosity is equal to the magnetic dipole luminosity at early times but becomes slightly lower at later times. The cases of SN 2018kyt and SN 2019hge are similar to that of PS1-12cil. The energy injected by the precessing magnetar into the supernova ejecta is given by
\begin{equation}
    E_{\rm in}=\int_{0}^{t_{\rm obs}}L_{\rm sd,pre}(t)dt,
\label{eq:E_in}
\end{equation}
where $t_{\rm obs}$ is the time of the last observational data point. In the case of homologous expansion, the kinetic energy of the supernova ejecta is given by \citep{Arnett1982}
\begin{equation}
    E_{\rm k}=\dfrac{3}{10}M_{\rm ej}v_{\rm ej}^{2}
\label{eq:E_k}
\end{equation}
By substituting the best-fit model parameters from Table \ref{tab:emcee_parameters} into the expression for the injected energy (equation \ref{eq:E_in}) and the expression for the ejecta kinetic energy (equation \ref{eq:E_k}), we find that, for all six SLSNe-I in our sample, the energy injected by the precessing magnetar into the supernova ejecta ($E_{\rm in}$) is smaller than the kinetic energy of the ejecta ($E_{\rm k}$). Taking SN 2019stc as an example, the ratio of injected energy to kinetic energy is $E_{\rm in}/E_{\rm k} = 0.128$. Therefore, the assumption adopted in our model that the supernova ejecta expand homologously with a constant velocity is reasonable. Similar treatments have also been adopted in previous studies, such as \cite{Inserra2013}, \cite{Wang2015_Ek}, and \cite{Wang2017}, in which this assumption is considered reasonable when the injected energy is smaller than the kinetic energy of the ejecta. 

Our work focuses on studying SLSNe-I by adopting the quasi-periodically varying magnetic dipole luminosity produced by a precessing magnetar as the input luminosity, which differs from the continuously decreasing magnetic dipole luminosity used in non-precessing magnetar-powered models. The primary difference between our precessing magnetar-powered model and the non-precessing magnetar-powered models is that the temporal evolution of the magnetic dipole luminosity used is different. On the other hand, the radiation mechanism and several assumptions, such as those regarding the thermalization efficiency and the ejecta expansion velocity, are consistent with those adopted in non-precessing magnetar-powered models.

As shown in Figure \ref{fig:emcee_lightcurve}, the theoretical light curves of SN 2019stc and SN 2021mkr continue to rise after the last observational data point, for the following reasons. The parameters used in the right panel of Figure \ref{fig:The_model_schematic_diagram} are the fitting results for SN 2019stc, which means that the right panel of Figure \ref{fig:The_model_schematic_diagram} presents the time evolution of $\alpha$, $L_{\rm sd,pre}$, and $L_{\rm mag}$ for SN 2019stc obtained from our theoretical model. As shown in Figure \ref{fig:emcee_lightcurve}, the last observational data point for SN 2019stc occurs at around 160 days. From the right panel of Figure \ref{fig:The_model_schematic_diagram}, it can be seen that at around 160 days, the magnetar in SN 2019stc is still undergoing precession, so $\alpha$ and $L_{\rm sd,pre}$ continue to exhibit quasi-periodic variations. Furthermore, $\alpha$ and $L_{\rm sd,pre}$ are in the initial rising phase around 160 days, and correspondingly, $L_{\rm mag}$ is also beginning to rise around 160 days. This explains why the theoretical light curve of SN 2019stc continues to rise after the last observational data point (around 160 days). Additionally, it is important to note that, as shown in the right panel of Figure \ref{fig:The_model_schematic_diagram}, both $L_{\rm sd,pre}$ and $L_{\rm mag}$ are decreasing, which implies that the theoretical light curve of SN 2019stc is overall decreasing. Furthermore, it can be seen that as the magnetar precession progresses, the rise in the theoretical light curve of SN 2019stc after the last data point will only last for a certain period, after which it will continue to decline. In other words, the last data point of SN 2019stc marks the beginning of the rising phase of the next bump, which has a lower luminosity and a longer duration. The reason for the continued rise of the theoretical light curve of SN 2021mkr after the last observation point is similar to that of SN 2019stc.

The variations in the energy injection rate caused by magnetar precession can only manifest as observable bumps in the light curves of SLSNe under specific circumstances. First, the precession period of the magnetar cannot be too short, otherwise the variations in the energy injection rate caused by magnetar precession will be completely smoothed out by the diffusion process of photons in the supernova ejecta, and will not be reflected in the light curves of SLSNe. Second, the precession period of the magnetar cannot exceed the observational timescale of the SLSNe, otherwise the variations in the energy injection rate caused by magnetar precession will not be observable within the observational window of the SLSNe. 

\section{Conclusion}
\label{sec:Con}
\cite{Hosseinzadeh2022} and \cite{Chen2023}, based on their studies of SLSNe-I samples, statistically concluded that SLSNe-I with post-peak bumps are common within the SLSNe-I population. Additionally, they proposed that models involving a time-varying central engine or multiple interactions between the ejecta and CSM can explain some of the SLSNe-I with post-peak bumps. On the other hand, in the theoretical explanations for the origin of post-peak bumps in SLSNe-I, some models have been proposed to explain a specific SLSN-I with post-peak bumps (e.g., \citealt{Fiore2021,Gomez2021,West2023}), while a few models have simultaneously provided theoretical explanations for multiple SLSNe-I with post-peak bumps (e.g., \citealt{DongXiao-Fei2023,ZhuJin-Ping2024}). Our study here belongs to the latter. Given that the magnetar-powered model is often considered one of the important energy mechanisms for SLSNe-I, we explore the origin of the post-peak bumps within the framework of the magnetar model. The precession of the magnetar can cause the energy injected into the supernova ejecta to deviate from the smooth energy injection predicted by the original magnetar-powered model, which may lead to the formation of post-peak bumps. Therefore, based on the original magnetar-powered model \citep{WangShan-Qin2015}, we consider the effects of magnetar precession and develop the precessing magnetar-powered model described in Section \ref{sec:precessing magnetar}.

In this work, we selected six representative SLSNe-I with one or more post-peak bumps from the literature to form our study sample. These six SLSNe-I are all considered by other studies to exhibit significant post-peak bumps (e.g., \citealt{Hosseinzadeh2022,Chen2023,DongXiao-Fei2023}) and have well-sampled multi-band photometric data. Their light curves exhibit complete rising and declining structures for both the primary peak and the post-peak bumps. To investigate whether the precessing magnetar-powered model introduced in Section \ref{sec:precessing magnetar} can explain the formation of post-peak bumps in SLSNe-I, we use the MCMC algorithm based on the emcee Python package \citep{Foreman-Mackey2013} to fit the multi-band light curves of the 6 SLSNe-I with post-peak bumps in our sample. This allows us to obtain the best-fit model parameters for each SLSNe-I (see Tabel \ref{tab:emcee_parameters}). We find that the multi-band photometric data of the 6 SLSNe-I with post-peak bumps in our sample can be well explained by our precessing magnetar-powered model (see Figure \ref{fig:emcee_lightcurve}). The model parameters also fall within reasonable ranges, indicating that our precessing magnetar-powered model is capable of explaining the observational data of SLSNe-I with one or more post-peak bumps. Therefore, our precessing magnetar-powered model provides a viable explanation for the observations of SLSNe-I with post-peak bumps.

By comparing the parameters of the 31 SLSNe-I obtained in \cite{YuYun-Wei2017}, we find that although the 6 SLSNe-I in our study exhibit post-peak bumps, their best-fit parameters are distributed within the typical range of SLSNe-I parameters (see figure \ref{fig:Our_VS_Yu2017}). This is consistent with the findings of previous studies (e.g., \citealt{Hosseinzadeh2022}). We conduct a comparison between the precessing magnetar parameters in SLSNe-I from our work and those in long GRBs from \cite{ZhangBiao2024} (see Figure \ref{fig:SLSNe_VS_GRBs}). We find that the ellipticity $\epsilon$ of precessing magnetars in SLSNe-I and long GRBs is correlated with the dipole magnetic field strength $B_{\rm p}$, while it appears to be unaffected by the initial spin period $P_0$. Combining the precessing magnetars of long GRBs and SLSNe-I, we find that the ellipticity $\epsilon$ and the dipole magnetic field strength $B_{\rm p}$ satisfy the relation given in equation (\ref{eq:log_epsilon_Bp}), with a correlation coefficient ${\rm r = 0.97}$. This finding may indicate that SLSNe-I and long GRBs share a common origin. Currently, the sample sizes of precessing magnetars in both SLSNe-I and long GRBs are limited, and therefore, our findings here remain to be verified. It would be meaningful to search for more SLSNe-I and long GRBs that support precessing magnetars as central engines, and to use a larger sample to validate our findings.

In addition to the main results discussed above, we have identified several other findings. The parameter $\delta$ in the magnetospheric factor expression (\ref{eq:lambda_alpha}), obtained by fitting the multi-band light curves of SLSNe-I in our sample, is distributed in the range of 0.79–1. This is consistent with the result of \cite{Spitkovsky2006}, who derived $\lambda(\alpha) \approx k_1 + k_2 \sin^2(\alpha)$ through numerical simulations, where $k_1 = 1 \pm 0.05$ and $k_2 = 1 \pm 0.11$. From equations (\ref{eq:dE_rot_dt}) and (\ref{eq:lambda_alpha}), or alternatively from equation (\ref{eq:Lsd_pre}) alone, it can be seen that $\delta$ affects the amplitude of the quasi-periodic variations in the spin-down luminosity generated by the precessing magnetar, thereby influencing the amplitude of the bumps in the resulting SLSNe light curves. The fitting results consistently yield $\delta \approx 1$, indicating that the amplitude of the bumps in the six SLSNe-I studied in our work is in good agreement with the bump amplitude predicted by a magnetar's magnetosphere being plasma-filled, where $\delta \approx 1$. The gravitational wave luminosity of a newborn magnetar relative to its electromagnetic radiation luminosity can be approximated as $L_{\rm GW}/L_{\rm EM} \sim 0.3(\epsilon/10^{-3})^2(B_{\rm p}/10^{15}{\rm G})^{-2}(P_0/{\rm ms})^{-2}$ \citep{Suvorov&Kokkotas2021}. By substituting the best-fit parameters of each SLSN-I in our sample into this expression, we find that all SLSNe-I in our sample satisfy $L_{\rm EM} \gg L_{\rm GW}$. This result is consistent with our assumption that the magnetar's spindown is dominated by magnetic dipole radiation.

Our work is the first to propose a precessing magnetar as the central engine for SLSNe-I and successfully explain the observational data of 6 representative SLSNe-I with post-peak bumps in our sample. This work provides further support for the magnetar as the central engine of SLSNe-I, and offers additional evidence for the early precession of newborn magnetars born from the collapse of massive stars. Our precessing magnetar-powered model has several advantages. On the one hand, compared to models that can only explain SLSNe-I with a single post-peak bump (e.g., \citealt{Chugai&Utrobin2022}), our model can account for both SLSNe-I with one post-peak bump and those with multiple post-peak bumps. On the other hand, some models explaining SLSNe-I with post-peak bumps require the introduction of additional energy sources, such as contributions from CSM interaction added to the magnetar-powered model \citep{Gomez2021}. In contrast, our model does not require any additional energy sources to explain SLSNe-I with post-peak bumps. It is worth noting that, as found in the studies by \cite{Hosseinzadeh2022} and \cite{Chen2023}, a subset of SLSNe-I with post-peak bumps cannot be explained by variations in the central engine. For these SLSNe-I, alternative models need to be considered, which may suggest that the post-peak bumps in SLSNe-I may have more than one origin. The origin of post-peak bumps in SLSNe-I remains an open question. Future observations and theoretical studies of post-peak bumps in more SLSNe-I are necessary and crucial to addressing this issue.

\section*{acknowledgments}
We thank the referee for valuable and helpful comments, which have helped us significantly improve our manuscript. We also thank Ji-Yu Cheng, Yu-Chen Huang, Zi-Bin Zhang, Jia-Pei Feng, and Min-Xuan Cai for useful discussions. This work was supported by the National Natural Science Foundation of China (grant No. 12393812) and the National SKA Program of China (grant No. 2020SKA0120302). L.L. is supported by the National Natural Science Foundation of China (grant No. 12303050). S.Q.Z. is supported by the starting Foundation of Guangxi University of Science and Technology (grant No. 24Z17).

\bibliographystyle{aasjournal}
\bibliography{refs}

\begin{thebibliography}{}
\expandafter\ifx\csname natexlab\endcsname\relax\def\natexlab#1{#1}\fi
\providecommand{\url}[1]{\href{#1}{#1}}
\providecommand{\dodoi}[1]{doi:~\href{http://doi.org/#1}{\nolinkurl{#1}}}
\providecommand{\doeprint}[1]{\href{http://ascl.net/#1}{\nolinkurl{http://ascl.net/#1}}}
\providecommand{\doarXiv}[1]{\href{https://arxiv.org/abs/#1}{\nolinkurl{https://arxiv.org/abs/#1}}}

\bibitem[{{Arnett}(1982)}]{Arnett1982}
{Arnett}, W.~D. 1982, \apj, 253, 785, \dodoi{10.1086/159681}

\bibitem[{{Arzamasskiy} {et~al.}(2015){Arzamasskiy}, {Philippov}, \&
  {Tchekhovskoy}}]{Arzamasskiy2015}
{Arzamasskiy}, L., {Philippov}, A., \& {Tchekhovskoy}, A. 2015, \mnras, 453,
  3540, \dodoi{10.1093/mnras/stv1818}

\bibitem[{{Chatzopoulos} {et~al.}(2012){Chatzopoulos}, {Wheeler}, \&
  {Vinko}}]{Chatzopoulos2012}
{Chatzopoulos}, E., {Wheeler}, J.~C., \& {Vinko}, J. 2012, \apj, 746, 121,
  \dodoi{10.1088/0004-637X/746/2/121}

\bibitem[{{Chatzopoulos} {et~al.}(2013){Chatzopoulos}, {Wheeler}, {Vinko},
  {Horvath}, \& {Nagy}}]{Chatzopoulos2013}
{Chatzopoulos}, E., {Wheeler}, J.~C., {Vinko}, J., {Horvath}, Z.~L., \& {Nagy},
  A. 2013, \apj, 773, 76, \dodoi{10.1088/0004-637X/773/1/76}

\bibitem[{{Chen} {et~al.}(2015){Chen}, {Smartt}, {Jerkstrand}, {Nicholl},
  {Bresolin}, {Kotak}, {Polshaw}, {Rest}, {Kudritzki}, {Zheng}, {Elias-Rosa},
  {Smith}, {Inserra}, {Wright}, {Kankare}, {Kangas}, \& {Fraser}}]{Chen2015}
{Chen}, T.~W., {Smartt}, S.~J., {Jerkstrand}, A., {et~al.} 2015, \mnras, 452,
  1567, \dodoi{10.1093/mnras/stv1360}

\bibitem[{{Chen} {et~al.}(2023{\natexlab{a}}){Chen}, {Yan}, {Kangas}, {Lunnan},
  {Sollerman}, {Schulze}, {Perley}, {Chen}, {Taggart}, {Hinds}, {Gal-Yam},
  {Wang}, {De}, {Bellm}, {Bloom}, {Dekany}, {Graham}, {Kasliwal}, {Kulkarni},
  {Laher}, {Neill}, \& {Rusholme}}]{Chen2023}
{Chen}, Z.~H., {Yan}, L., {Kangas}, T., {et~al.} 2023{\natexlab{a}}, \apj, 943,
  42, \dodoi{10.3847/1538-4357/aca162}

\bibitem[{{Chen} {et~al.}(2023{\natexlab{b}}){Chen}, {Yan}, {Kangas}, {Lunnan},
  {Schulze}, {Sollerman}, {Perley}, {Chen}, {Taggart}, {Hinds}, {Gal-Yam},
  {Wang}, {Andreoni}, {Bellm}, {Bloom}, {Burdge}, {Burgos}, {Cook}, {Dahiwale},
  {De}, {Dekany}, {Dugas}, {Frederik}, {Fremling}, {Graham}, {Hankins}, {Ho},
  {Jencson}, {Karambelkar}, {Kasliwal}, {Kulkarni}, {Laher}, {Rusholme},
  {Sharma}, {Taddia}, {Tartaglia}, {Thomas}, {Tzanidakis}, {Van Roestel},
  {Walter}, {Yang}, {Yao}, \& {Yaron}}]{Chen2023_photometry}
---. 2023{\natexlab{b}}, \apj, 943, 41, \dodoi{10.3847/1538-4357/aca161}

\bibitem[{{Chevalier} \& {Irwin}(2011)}]{Chevalier&Irwin2011}
{Chevalier}, R.~A., \& {Irwin}, C.~M. 2011, \apjl, 729, L6,
  \dodoi{10.1088/2041-8205/729/1/L6}

\bibitem[{{Chugai} \& {Utrobin}(2022)}]{Chugai&Utrobin2022}
{Chugai}, N.~N., \& {Utrobin}, V.~P. 2022, \mnras, 512, L71,
  \dodoi{10.1093/mnrasl/slab131}

\bibitem[{{Dessart} {et~al.}(2012){Dessart}, {Hillier}, {Waldman}, {Livne}, \&
  {Blondin}}]{Dessart2012}
{Dessart}, L., {Hillier}, D.~J., {Waldman}, R., {Livne}, E., \& {Blondin}, S.
  2012, \mnras, 426, L76, \dodoi{10.1111/j.1745-3933.2012.01329.x}

\bibitem[{{Dong} {et~al.}(2023){Dong}, {Liu}, {Gao}, \&
  {Yang}}]{DongXiao-Fei2023}
{Dong}, X.-F., {Liu}, L.-D., {Gao}, H., \& {Yang}, S. 2023, \apj, 951, 61,
  \dodoi{10.3847/1538-4357/acd848}

\bibitem[{{Fiore} {et~al.}(2021){Fiore}, {Chen}, {Jerkstrand}, {Benetti},
  {Ciolfi}, {Inserra}, {Cappellaro}, {Pastorello}, {Leloudas}, {Schulze},
  {Berton}, {Burke}, {McCully}, {Fong}, {Galbany}, {Gromadzki},
  {Guti{\'e}rrez}, {Hiramatsu}, {Hosseinzadeh}, {Howell}, {Kankare}, {Lunnan},
  {M{\"u}ller-Bravo}, {O'Neill}, {Nicholl}, {Rau}, {Sollerman}, {Terreran},
  {Valenti}, \& {Young}}]{Fiore2021}
{Fiore}, A., {Chen}, T.~W., {Jerkstrand}, A., {et~al.} 2021, \mnras, 502, 2120,
  \dodoi{10.1093/mnras/staa4035}

\bibitem[{{Foreman-Mackey} {et~al.}(2013){Foreman-Mackey}, {Hogg}, {Lang}, \&
  {Goodman}}]{Foreman-Mackey2013}
{Foreman-Mackey}, D., {Hogg}, D.~W., {Lang}, D., \& {Goodman}, J. 2013, \pasp,
  125, 306, \dodoi{10.1086/670067}

\bibitem[{{Fremling} {et~al.}(2020){Fremling}, {Miller}, {Sharma}, {Dugas},
  {Perley}, {Taggart}, {Sollerman}, {Goobar}, {Graham}, {Neill}, {Nordin},
  {Rigault}, {Walters}, {Andreoni}, {Bagdasaryan}, {Belicki}, {Cannella},
  {Bellm}, {Cenko}, {De}, {Dekany}, {Frederick}, {Golkhou}, {Graham}, {Helou},
  {Ho}, {Kasliwal}, {Kupfer}, {Laher}, {Mahabal}, {Masci}, {Riddle},
  {Rusholme}, {Schulze}, {Shupe}, {Smith}, {van Velzen}, {Yan}, {Yao},
  {Zhuang}, \& {Kulkarni}}]{Fremling2020}
{Fremling}, C., {Miller}, A.~A., {Sharma}, Y., {et~al.} 2020, \apj, 895, 32,
  \dodoi{10.3847/1538-4357/ab8943}

\bibitem[{{Gaensler} \& {Slane}(2006)}]{Gaensler&Slane2006}
{Gaensler}, B.~M., \& {Slane}, P.~O. 2006, \araa, 44, 17,
  \dodoi{10.1146/annurev.astro.44.051905.092528}

\bibitem[{{Gal-Yam}(2012)}]{Gal-Yam2012}
{Gal-Yam}, A. 2012, Science, 337, 927, \dodoi{10.1126/science.1203601}

\bibitem[{{Gal-Yam}(2019)}]{Gal-Yam2019}
---. 2019, \araa, 57, 305, \dodoi{10.1146/annurev-astro-081817-051819}

\bibitem[{{Gao} {et~al.}(2023){Gao}, {Shao}, {Desvignes}, {Jones}, {Kramer}, \&
  {Yim}}]{GaoYong2023}
{Gao}, Y., {Shao}, L., {Desvignes}, G., {et~al.} 2023, \mnras, 519, 1080,
  \dodoi{10.1093/mnras/stac3546}

\bibitem[{{Ginzburg} \& {Balberg}(2012)}]{Ginzburg&Balberg2012}
{Ginzburg}, S., \& {Balberg}, S. 2012, \apj, 757, 178,
  \dodoi{10.1088/0004-637X/757/2/178}

\bibitem[{{Goldreich} \& {Julian}(1969)}]{Goldreich&Julian1969}
{Goldreich}, P., \& {Julian}, W.~H. 1969, \apj, 157, 869,
  \dodoi{10.1086/150119}

\bibitem[{{Gomez} {et~al.}(2021){Gomez}, {Berger}, {Hosseinzadeh}, {Blanchard},
  {Nicholl}, \& {Villar}}]{Gomez2021}
{Gomez}, S., {Berger}, E., {Hosseinzadeh}, G., {et~al.} 2021, \apj, 913, 143,
  \dodoi{10.3847/1538-4357/abf5e3}

\bibitem[{{Hosseinzadeh} {et~al.}(2022){Hosseinzadeh}, {Berger}, {Metzger},
  {Gomez}, {Nicholl}, \& {Blanchard}}]{Hosseinzadeh2022}
{Hosseinzadeh}, G., {Berger}, E., {Metzger}, B.~D., {et~al.} 2022, \apj, 933,
  14, \dodoi{10.3847/1538-4357/ac67dd}

\bibitem[{{Inserra} {et~al.}(2013){Inserra}, {Smartt}, {Jerkstrand}, {Valenti},
  {Fraser}, {Wright}, {Smith}, {Chen}, {Kotak}, {Pastorello}, {Nicholl},
  {Bresolin}, {Kudritzki}, {Benetti}, {Botticella}, {Burgett}, {Chambers},
  {Ergon}, {Flewelling}, {Fynbo}, {Geier}, {Hodapp}, {Howell}, {Huber},
  {Kaiser}, {Leloudas}, {Magill}, {Magnier}, {McCrum}, {Metcalfe}, {Price},
  {Rest}, {Sollerman}, {Sweeney}, {Taddia}, {Taubenberger}, {Tonry},
  {Wainscoat}, {Waters}, \& {Young}}]{Inserra2013}
{Inserra}, C., {Smartt}, S.~J., {Jerkstrand}, A., {et~al.} 2013, \apj, 770,
  128, \dodoi{10.1088/0004-637X/770/2/128}

\bibitem[{{Inserra} {et~al.}(2017){Inserra}, {Nicholl}, {Chen}, {Jerkstrand},
  {Smartt}, {Kr{\"u}hler}, {Anderson}, {Baltay}, {Della Valle}, {Fraser},
  {Gal-Yam}, {Galbany}, {Kankare}, {Maguire}, {Rabinowitz}, {Smith}, {Valenti},
  \& {Young}}]{Inserra2017}
{Inserra}, C., {Nicholl}, M., {Chen}, T.~W., {et~al.} 2017, \mnras, 468, 4642,
  \dodoi{10.1093/mnras/stx834}

\bibitem[{{Kalapotharakos} \&
  {Contopoulos}(2009)}]{Kalapotharakos&Contopoulos2009}
{Kalapotharakos}, C., \& {Contopoulos}, I. 2009, \aap, 496, 495,
  \dodoi{10.1051/0004-6361:200810281}

\bibitem[{{Kasen} \& {Bildsten}(2010)}]{Kasen&Bildsten2010}
{Kasen}, D., \& {Bildsten}, L. 2010, \apj, 717, 245,
  \dodoi{10.1088/0004-637X/717/1/245}

\bibitem[{{Kasen} {et~al.}(2016){Kasen}, {Metzger}, \& {Bildsten}}]{Kasen2016}
{Kasen}, D., {Metzger}, B.~D., \& {Bildsten}, L. 2016, \apj, 821, 36,
  \dodoi{10.3847/0004-637X/821/1/36}

\bibitem[{{Kotera} {et~al.}(2013){Kotera}, {Phinney}, \& {Olinto}}]{Kotera2013}
{Kotera}, K., {Phinney}, E.~S., \& {Olinto}, A.~V. 2013, \mnras, 432, 3228,
  \dodoi{10.1093/mnras/stt680}

\bibitem[{{Levin} {et~al.}(2020){Levin}, {Beloborodov}, \&
  {Bransgrove}}]{Levin2020}
{Levin}, Y., {Beloborodov}, A.~M., \& {Bransgrove}, A. 2020, \apjl, 895, L30,
  \dodoi{10.3847/2041-8213/ab8c4c}

\bibitem[{{Li} {et~al.}(2020){Li}, {Wang}, {Liu}, {Wang}, {Liang}, \&
  {Dai}}]{LiLong2020}
{Li}, L., {Wang}, S.-Q., {Liu}, L.-D., {et~al.} 2020, \apj, 891, 98,
  \dodoi{10.3847/1538-4357/ab718d}

\bibitem[{{Li} {et~al.}(2024){Li}, {Zhong}, {Xiao}, {Dai}, {Huang}, \&
  {Sheng}}]{Li2024}
{Li}, L., {Zhong}, S.-Q., {Xiao}, D., {et~al.} 2024, \apjl, 963, L13,
  \dodoi{10.3847/2041-8213/ad2611}

\bibitem[{{Liu} {et~al.}(2018){Liu}, {Wang}, {Wang}, \&
  {Dai}}]{LiuLiang-Duan2018}
{Liu}, L.-D., {Wang}, L.-J., {Wang}, S.-Q., \& {Dai}, Z.-G. 2018, \apj, 856,
  59, \dodoi{10.3847/1538-4357/aab157}

\bibitem[{{Liu} {et~al.}(2017){Liu}, {Wang}, {Wang}, {Dai}, {Yu}, \&
  {Peng}}]{LiuLiang-Duan2017}
{Liu}, L.-D., {Wang}, S.-Q., {Wang}, L.-J., {et~al.} 2017, \apj, 842, 26,
  \dodoi{10.3847/1538-4357/aa73d9}

\bibitem[{{Lunnan} {et~al.}(2018){Lunnan}, {Chornock}, {Berger}, {Jones},
  {Rest}, {Czekala}, {Dittmann}, {Drout}, {Foley}, {Fong}, {Kirshner},
  {Laskar}, {Leibler}, {Margutti}, {Milisavljevic}, {Narayan}, {Pan}, {Riess},
  {Roth}, {Sanders}, {Scolnic}, {Smartt}, {Smith}, {Chambers}, {Draper},
  {Flewelling}, {Huber}, {Kaiser}, {Kudritzki}, {Magnier}, {Metcalfe},
  {Wainscoat}, {Waters}, \& {Willman}}]{Lunnan2018}
{Lunnan}, R., {Chornock}, R., {Berger}, E., {et~al.} 2018, \apj, 852, 81,
  \dodoi{10.3847/1538-4357/aa9f1a}

\bibitem[{{Makishima} {et~al.}(2014){Makishima}, {Enoto}, {Hiraga}, {Nakano},
  {Nakazawa}, {Sakurai}, {Sasano}, \& {Murakami}}]{Makishima2014}
{Makishima}, K., {Enoto}, T., {Hiraga}, J.~S., {et~al.} 2014, \prl, 112,
  171102, \dodoi{10.1103/PhysRevLett.112.171102}

\bibitem[{{Metzger} {et~al.}(2015){Metzger}, {Margalit}, {Kasen}, \&
  {Quataert}}]{Metzger2015}
{Metzger}, B.~D., {Margalit}, B., {Kasen}, D., \& {Quataert}, E. 2015, \mnras,
  454, 3311, \dodoi{10.1093/mnras/stv2224}

\bibitem[{{Michel} \& {Goldwire}(1970)}]{Michel&Goldwire1970}
{Michel}, F.~C., \& {Goldwire}, H.~C., J. 1970, \aplett, 5, 21

\bibitem[{{Moriya} {et~al.}(2022){Moriya}, {Murase}, {Kashiyama}, \&
  {Blinnikov}}]{Moriya2022}
{Moriya}, T.~J., {Murase}, K., {Kashiyama}, K., \& {Blinnikov}, S.~I. 2022,
  \mnras, 513, 6210, \dodoi{10.1093/mnras/stac1352}

\bibitem[{{Moriya} {et~al.}(2018){Moriya}, {Sorokina}, \&
  {Chevalier}}]{Moriya2018}
{Moriya}, T.~J., {Sorokina}, E.~I., \& {Chevalier}, R.~A. 2018, \ssr, 214, 59,
  \dodoi{10.1007/s11214-018-0493-6}

\bibitem[{{Nicholl} {et~al.}(2017){Nicholl}, {Guillochon}, \&
  {Berger}}]{Nicholl2017}
{Nicholl}, M., {Guillochon}, J., \& {Berger}, E. 2017, \apj, 850, 55,
  \dodoi{10.3847/1538-4357/aa9334}

\bibitem[{{Ostriker} \& {Gunn}(1969)}]{Ostriker&Gunn1969}
{Ostriker}, J.~P., \& {Gunn}, J.~E. 1969, \apj, 157, 1395,
  \dodoi{10.1086/150160}

\bibitem[{{Perley} {et~al.}(2020){Perley}, {Fremling}, {Sollerman}, {Miller},
  {Dahiwale}, {Sharma}, {Bellm}, {Biswas}, {Brink}, {Bruch}, {De}, {Dekany},
  {Drake}, {Duev}, {Filippenko}, {Gal-Yam}, {Goobar}, {Graham}, {Graham}, {Ho},
  {Irani}, {Kasliwal}, {Kim}, {Kulkarni}, {Mahabal}, {Masci}, {Modak}, {Neill},
  {Nordin}, {Riddle}, {Soumagnac}, {Strotjohann}, {Schulze}, {Taggart},
  {Tzanidakis}, {Walters}, \& {Yan}}]{Perley2020}
{Perley}, D.~A., {Fremling}, C., {Sollerman}, J., {et~al.} 2020, \apj, 904, 35,
  \dodoi{10.3847/1538-4357/abbd98}

\bibitem[{{Philippov} {et~al.}(2015){Philippov}, {Spitkovsky}, \&
  {Cerutti}}]{Philippov2015}
{Philippov}, A.~A., {Spitkovsky}, A., \& {Cerutti}, B. 2015, \apjl, 801, L19,
  \dodoi{10.1088/2041-8205/801/1/L19}

\bibitem[{{Smith} \& {McCray}(2007)}]{Smith&McCray2007}
{Smith}, N., \& {McCray}, R. 2007, \apjl, 671, L17, \dodoi{10.1086/524681}

\bibitem[{{Spitkovsky}(2006)}]{Spitkovsky2006}
{Spitkovsky}, A. 2006, \apjl, 648, L51, \dodoi{10.1086/507518}

\bibitem[{{Suvorov} \& {Kokkotas}(2020)}]{Suvorov&Kokkotas2020}
{Suvorov}, A.~G., \& {Kokkotas}, K.~D. 2020, \apjl, 892, L34,
  \dodoi{10.3847/2041-8213/ab8296}

\bibitem[{{Suvorov} \& {Kokkotas}(2021)}]{Suvorov&Kokkotas2021}
---. 2021, \mnras, 502, 2482, \dodoi{10.1093/mnras/stab153}

\bibitem[{{Umeda} \& {Nomoto}(2008)}]{Umeda&Nomoto2008}
{Umeda}, H., \& {Nomoto}, K. 2008, \apj, 673, 1014, \dodoi{10.1086/524767}

\bibitem[{{Vurm} \& {Metzger}(2021)}]{Vurm2021}
{Vurm}, I., \& {Metzger}, B.~D. 2021, \apj, 917, 77,
  \dodoi{10.3847/1538-4357/ac0826}

\bibitem[{{Wang} {et~al.}(2017){Wang}, {Cano}, {Wang}, {Zheng}, {Dai},
  {Filippenko}, \& {Liu}}]{Wang2017}
{Wang}, S.-Q., {Cano}, Z., {Wang}, L.-J., {et~al.} 2017, \apj, 850, 148,
  \dodoi{10.3847/1538-4357/aa95c5}

\bibitem[{{Wang} {et~al.}(2019){Wang}, {Wang}, \& {Dai}}]{WangShan-Qin2019}
{Wang}, S.-Q., {Wang}, L.-J., \& {Dai}, Z.-G. 2019, Research in Astronomy and
  Astrophysics, 19, 063, \dodoi{10.1088/1674-4527/19/5/63}

\bibitem[{{Wang} {et~al.}(2015{\natexlab{a}}){Wang}, {Wang}, {Dai}, \&
  {Wu}}]{WangShan-Qin2015}
{Wang}, S.~Q., {Wang}, L.~J., {Dai}, Z.~G., \& {Wu}, X.~F. 2015{\natexlab{a}},
  \apj, 799, 107, \dodoi{10.1088/0004-637X/799/1/107}

\bibitem[{{Wang} {et~al.}(2015{\natexlab{b}}){Wang}, {Wang}, {Dai}, \&
  {Wu}}]{Wang2015_Ek}
---. 2015{\natexlab{b}}, \apj, 807, 147, \dodoi{10.1088/0004-637X/807/2/147}

\bibitem[{{Wei} {et~al.}(2022){Wei}, {Zhao}, \& {Wang}}]{WeiYu-Jia2022}
{Wei}, Y.-J., {Zhao}, Z.-Y., \& {Wang}, F.-Y. 2022, \aap, 658, A163,
  \dodoi{10.1051/0004-6361/202142321}

\bibitem[{{West} {et~al.}(2023){West}, {Lunnan}, {Omand}, {Kangas}, {Schulze},
  {Strotjohann}, {Yang}, {Fransson}, {Sollerman}, {Perley}, {Yan}, {Chen},
  {Chen}, {Taggart}, {Fremling}, {Bloom}, {Drake}, {Graham}, {Kasliwal},
  {Laher}, {Medford}, {Neill}, {Riddle}, \& {Shupe}}]{West2023}
{West}, S.~L., {Lunnan}, R., {Omand}, C.~M.~B., {et~al.} 2023, \aap, 670, A7,
  \dodoi{10.1051/0004-6361/202244086}

\bibitem[{{Woosley}(2010)}]{Woosley2010}
{Woosley}, S.~E. 2010, \apjl, 719, L204, \dodoi{10.1088/2041-8205/719/2/L204}

\bibitem[{{Yang} \& {Dai}(2019)}]{Yang&Dai2019}
{Yang}, Y.-H., \& {Dai}, Z.-G. 2019, \apj, 885, 149,
  \dodoi{10.3847/1538-4357/ab48dd}

\bibitem[{{Yu} \& {Li}(2017)}]{YuYun-Wei&Li2017}
{Yu}, Y.-W., \& {Li}, S.-Z. 2017, \mnras, 470, 197,
  \dodoi{10.1093/mnras/stx1028}

\bibitem[{{Yu} {et~al.}(2017){Yu}, {Zhu}, {Li}, {L{\"u}}, \&
  {Zou}}]{YuYun-Wei2017}
{Yu}, Y.-W., {Zhu}, J.-P., {Li}, S.-Z., {L{\"u}}, H.-J., \& {Zou}, Y.-C. 2017,
  \apj, 840, 12, \dodoi{10.3847/1538-4357/aa6c27}

\bibitem[{{Zanazzi} \& {Lai}(2015)}]{Zanazzi&Lai2015}
{Zanazzi}, J.~J., \& {Lai}, D. 2015, \mnras, 451, 695,
  \dodoi{10.1093/mnras/stv955}

\bibitem[{{Zanazzi} \& {Lai}(2020)}]{Zanazzi&Lai2020}
---. 2020, \apjl, 892, L15, \dodoi{10.3847/2041-8213/ab7cdd}

\bibitem[{{Zhang} {et~al.}(2024){Zhang}, {Zhong}, {Li}, \&
  {Dai}}]{ZhangBiao2024}
{Zhang}, B., {Zhong}, S.-Q., {Li}, L., \& {Dai}, Z.-G. 2024, \apj, 977, 206,
  \dodoi{10.3847/1538-4357/ad9005}

\bibitem[{{Zhu} {et~al.}(2024){Zhu}, {Liu}, {Yu}, {Mandel}, {Hirai}, {Zhang},
  \& {Chen}}]{ZhuJin-Ping2024}
{Zhu}, J.-P., {Liu}, L.-D., {Yu}, Y.-W., {et~al.} 2024, \apjl, 970, L42,
  \dodoi{10.3847/2041-8213/ad63a8}

\bibitem[{{Zou} \& {Liang}(2022)}]{Zou&Liang2022}
{Zou}, L., \& {Liang}, E.-W. 2022, \mnras, 513, L89,
  \dodoi{10.1093/mnrasl/slac040}

\end{thebibliography}

\end{document}